\documentclass[twoside,english,sort&compress]{iopart}
\pdfoutput=1
\usepackage[latin9]{inputenc}
\usepackage{geometry}
\usepackage{textcomp}
\usepackage{tipa}
\usepackage{tipx}
\usepackage{amsbsy}
\usepackage{amstext}
\usepackage{amsthm}
\usepackage{graphicx}
\usepackage[numbers]{natbib}

\makeatletter
\usepackage{iopams}
\usepackage{setstack}
\theoremstyle{plain}
\newtheorem{thm}{\protect\theoremname}
  \theoremstyle{remark}
  \newtheorem{rem}[thm]{\protect\remarkname}

\newcommand{\eqref}[1]{(\ref{#1})}

\makeatother

\usepackage{babel}
  \providecommand{\remarkname}{Remark}
\providecommand{\theoremname}{Theorem}

\begin{document}

\title[Large deviations in RPM]{Large deviations of avalanches in the raise and peel model}

\author{{A.M. Povolotsky$^{1,2,*}$, P. Pyatov$^{2,1,\dagger}$,
V. Rittenberg}$^{3,\ddagger}$}

\address{{$^{1}$Bogoliubov Laboratory of Theoretical Physics, Joint
Institute for Nuclear Research, 141980, Dubna, Russia}\\
{$^{2}$National Research University Higher School of Economics,
20 Myasnitskaya, 101000, Moscow, Russia}\\
{$^{3}$Physikalisches Institut, Universit\"{ }at Bonn, Nussallee
12, 53115 Bonn, Germany}}

\ead{$^{*}$alexander.povolotsky@gmail.com,$^{\dagger}$pyatov@theor.jinr.ru,$^{\ddagger}$vladimir@th.physik.uni-bonn.de}
\begin{abstract}
We study the large deviation functions for two quantities characterizing
the avalanche dynamics in the Raise and Peel model: the number of
tiles removed by avalanches and the number of global avalanches extending
through the whole system. To this end, we exploit their connection
to the groundstate eigenvalue of the XXZ model with twisted boundary
conditions. We evaluate the cumulants of the two quantities asymptotically
in the limit of the large system size. The first cumulants, the means,
confirm the exact formulas conjectured from analysis of finite systems.
We discuss the phase transition from critical to non-critical behaviour
in the rate function of the global avalanches conditioned to an atypical
values of the number of tiles removed by avalanches per unit time.
\end{abstract}




\maketitle

\section{Introduction}

The large deviation theory is a way to bring  concepts from the
equilibrium thermodynamics to the non-equilibrium context. In particular,
it is well known that in a large system in equilibrium with environment
the probability distribution of a macroscopic additive quantity $X$,
e.g. energy, number of particles or charge, is an exponential of the
entropy, $P(X\approx x)\asymp\exp S(x)$, where the latter is normally
an extensive function of $x$, i.e. $S(x)\sim Vs(x/V)$ as $V\to\infty$,
where $V$ is a volume or any other additive coordinate. Below in the
equations like this we  use the notation $\approx$ to state
that $X$ is in a small vicinity of $x$ and the sign $\asymp$ means
that the ratio of logarithms of both sides tend to one as $x\to\infty$.

If we think about the time evolution of, to be specific, a stochastic
system and measure an additive functional $Y_{t}$ over its trajectories,
it is a common situation that at large time $t\to\infty$ the probability
distribution of $Y_{t}$ takes the form
\[
P(Y_{t}/t\approx y)\asymp\exp\left(-tI(y)\right).
\]
This is to say that $Y_{t}$ satisfies the large deviation property
with the rate function $I(y).$ Therefore, the rate function contains
exactly the same information about the space-time ensemble of trajectories
of a stochastic process as the specific entropy about an equilibrium
system.

Alternative description of the equilibrium state can be given in terms
of the Legendre transform of the entropy, the free energy $f(\mu)=-\sup_{x}(s(x)+\mu x)$,
which also can be viewed as the generating function of scaled cumulants
of $X.$  Similarly, the Legendre transform applied to the rate function
$I(y)$ yields the scaled cumulant generating function (CGF)
\[
\hat{I}(\gamma)=\sup_{y}\left(y\gamma-I(y)\right)=\lim_{t\to\infty}t^{-1}\log\mathbb{E}\left(e^{\gamma Y_{t}}\right).
\]
It gives an equivalent description of time history of the system when
$I(y)$ is convex.

Having its roots in Boltzmann's discovery of statistical nature of
the entropy the large deviations theory was first pioneered by Cram\`er \cite{Cramer}
with application to sums of independent identically distributed random
variables and then developed by Freidlin, Wentzell \cite{FW} for stochastic
dynamical systems  and by Donsker and Varadhan \cite{DV1}-\cite{DV3} for Markov processes.

One of the major applications of the theory became the stochastic
systems of interacting particles, which served as toy models of non-equilibrium
systems. Recently a remarkable progress was achieved for several exactly
solvable models of interacting particles (for a review  see \cite{Derrida_Rev} and references therein). Formulated in terms of Markov
processes they admitted a construction of the stationary measure or
even a full diagonalization of the Markov matrix by means of the matrix
product ansatz or the Bethe ansatz, respectively. In this way the
large deviation functions (LDF) of particle density and particle current
were obtained. They served as first examples of LDF obtained for
the diffusive and driven diffusive systems. Appearance of a number
of exact results eventually culminated in a formulation of the macroscopic
fluctuation theory, a kind of universal field theory approach to large
deviations in general diffusive systems, which is not restricted solely
to exactly solvable cases.

Interestingly, in some cases the finite size corrections to the scaled CGFs of particle
current were conjectured to extend beyond the realm of integrable
models \cite{DL},\cite{ADLF}. This is to say, that their functional forms are universal
within the whole Edwards-Wilkinson (EW) \cite{EW}  and Kardar-Parisi-Zhang (KPZ) \cite{KPZ}
universality classes, respectively. As usual, the nature of universality
depends on the symmetries or conservation laws, rather than on microscopic
details of the model. In the mentioned cases it is a single hyperbolic
conservation law responsible the Eulerian hydrodynamics (linear or nonlinear
for EW and KPZ  classes respectively). This remark is to emphasize
that often the LDF is  not only a characteristic of a particular
model, but rather has a universal meaning within a much wider context.

The present work is devoted to large deviations in the Raise and Peel
model (RPM), which is a stochastic model possessing yet another class
of symmetries, very different from the previously mentioned cases,
the conformal symmetry. The model was proposed in \citep{GNPR} as
a model of interface, which moves up by local random deposition of
tiles onto a substrate and moves down by spontaneous non-local avalanche-like
evaporation of tiles. It was formulated in the language of representations
of Temperley-Lieb (TL) algebra and dynamical rules of the model were
dictated by the generator relations. It attracted significant attention
due to its reach combinatorial content.

Initially the studies were concentrated on the stationary state of
the process. A number of exact results on the structure of the stationary
state were obtained and even more conjectures were proposed by exploiting
connections between RPM and the six vertex model, XXZ model,  the
fully packed loop model, the O(1) loop model and alternating sign
matrices (see \citep{AR-07,G} and references therein). Some other
predictions on space and time correlation functions in RPM were made
with the use of conformal field theory \citep{AR-15}.

Thus, most of the results on RPM obtained to date concern the properties
of the stationary state, while almost no any information involving
time-time correlations is yet available. In contrast, in the present
paper we study the time evolution of the model, concentrating on the
characteristics of the avalanche dynamics. We study the joint large deviations
of two time integrated quantities: the number $\mathcal{N}_{t}^{\lozenge}$
of all tiles removed within the avalanches by given time $t$ and
the total number $\mathcal{N}_{t}^{\circlearrowleft}$ of global avalanches,
those which extend over the whole system. Using the mapping
of the RPM to the XXZ quantum chain we obtain the largest eigenvalue of
Markov matrix, deformed by inclusion of two parameters responsible
for counting the mentioned quantities. This deformation gives the
meaning of the generating function of joint scaled cumulants of $\mathcal{N}_{t}^{\lozenge}$
and $\mathcal{N}_{t}^{\circlearrowleft}$ to the largest eigenvalue.
By applying the Legendre transform to this function we obtain the
joint rate function for the two quantities. The result is obtained
in the thermodynamic limit in two leading orders in the system size.
The leading order, the thermodynamic value of the eigenvalue obtained,
is responsible for the distribution of $\mathcal{N}_{t}^{\lozenge}$
in the thermodynamic limit. Similarly to the thermodynamic value of
free energy of two-dimensional vertex models, this quantity is specific
for the particular model. The next to the leading finite size correction
contains information about the distribution of $\mathcal{N}_{t}^{\circlearrowleft}$,
the corrections to the distribution of $\mathcal{N}_{t}^{\lozenge}$
and the mutual dependence of the two quantities. Unlike the
leading order, this correction is predicted by the conformal field
theory, and thus is expected to be universal to some extent.
Our result allow us to obtain the exact (within the two leading orders)
cumulants of both $\mathcal{N}_{t}^{\lozenge}$ and $\mathcal{N}_{t}^{\circlearrowleft}$.
The cumulants of the first order, the mean, can also be obtained from
averaging over the stationary states. Our results confirm the finite
size formulas conjectured from the analysis of the stationary states
of finite systems. The higher order cumulants are purely dynamical
quantities, which are obtained first time in the present paper. We
also study the asymptotics of the rate function obtained and discuss
the phase transition from critical to non-critical phase observed
in the large deviation functions of $\mathcal{N}_{t}^{\circlearrowleft}$
conditioned to an atypical value of the time average $\mathcal{N}_{t}^{\lozenge}/t$.

The article is organized as follows. In Section 2 we remind the basics
about the Temeperley-Lieb algebra and the Periodic Temperley-Lieb
algebra and formulate the Raise and Peel model defining its Markov
generator in terms of the generators of the Periodic Temperley-Lieb
algebra. Then we describe the R-matrix representations of the Periodic
Temperley-Lieb algebra, rewrite the stochastic generator in terms
of XXZ Hamiltonian with the twist and anisotropy parameters fixed
at stochastic point and describe its diagonalization by the Bethe
ansatz. In the last subsection of this section we discuss the stationary
state of the Raise and Peel model and state two conjectures about
the time averages of $\mathcal{N}_{t}^{\lozenge}$ and $\mathcal{N}_{t}^{\circlearrowleft}$.
In Section 3 we introduce a non-stochastic deformation of the Markov
generator, which describes an evolution of the joint moment generating
function of the quantities of interest, and relate it to the Hamiltonian
of XXZ chain with the twist and anisotropy parameters taking generic
values. We survey the formulas for the groundstate eigenvalue of the
XXZ Hamiltonian in the thermodynamic limit and for the finite size corrections
to it and then reinterpret them in terms of scaled cumulant generating
function of  $\mathcal{N}_{t}^{\lozenge}$ and $\mathcal{N}_{t}^{\circlearrowleft}$.
The end of this section contains main results of the article. In Section
4 we discuss and interpret the results obtained and mention some unsolved
problems. The explicit values and asymptotics of integrals and sums
used for our analysis are listed in  \ref{sec:Particular-values-of-Y}.

\section{Raise and Peel model, Temperley-Lieb algebra and XXZ chain}

\subsection{Temperley-Lieb algebra}

The dynamical rules of the RPM are imposed by a structure of the Temperley-Lieb
(TL) algebra. In the present paper we deal with the RPM on the periodic
lattice of even size $L$. It is defined in terms of the periodic
TL algebra $PTL_{L}$ \citep{L} generated by $L$ operators $e_{1},\dots,e_{L}$
satisfying the usual TL relations
\begin{eqnarray}
e_{i}^{2} & = & 2_{q}e_{i},\label{eq: TL1}\\
e_{i}e_{i\pm1}e_{i} & = & e_{i},\label{eq: TL2}\\
e_{i}e_{j} & = & e_{j}e_{i},\quad\left|i-j\right|>1,\label{eq: TL3}
\end{eqnarray}
where $i,j=1,\dots,L\,\,(\mbox{mod}\,\,L)$, which means that we impose
periodicity condition $e_{L+1}\equiv e_{1}$. The coefficient in (\ref{eq: TL1})
is the algebra coefficient $2_{q}=q+q^{-1}$ parametrized by $q\in\mathbb{C}\setminus\{0\}$.
The periodic Temperley-Lieb algebra is infinite dimensional. For our
purposes we need its (still infinite dimensional) quotient algebra
by relations
\begin{equation}
J_{L}I_{L}J_{L}=\kappa J_{L},\qquad I_{L}J_{L}I_{L}=\kappa I_{L},\label{eq: TL_periodic}
\end{equation}
where
\begin{equation}
I_{L}=e_{2}e_{4}\dots e_{L},\qquad J_{L}=e_{1}e_{3}\dots e_{L-1}\label{eq: I,J}
\end{equation}
are the (unnormalized) projectors, and $\kappa\in\mathbb{C}\setminus\{0\}$
is yet another algebra parameter.

In the following we consider a Markov process on a finite dimensional
state space with states represented by monomials from the left ideal
$\mathcal{J}_{L}$ of the quotient algebra (\ref{eq: TL1}\textendash \ref{eq: TL_periodic})
generated by $J_{L}$ (equivalently one can consider the ideal $\mathcal{I}_{L}$
generated by $I_{L}$). Specifically, we choose the monomial $J_{L}$
from (\ref{eq: I,J}) to be an initial state. We call it \textit{substrate}.
The other states are obtained by any sequence of the generators acting on $J_{L}$ from the left.

Two different states are represented by two nonequivalent monomials,
where by nonequivalent we mean that they can not be reduced to one
another by applying relations (\ref{eq: TL1}-\ref{eq: TL_periodic}).
Altogether one finds ${L \choose L/2}$ states in the ideal $\mathcal{J}_{L}$.
Note that an application of relations (\ref{eq: TL1}), (\ref{eq: TL_periodic})
also yields the numerical coefficients, which are the integer powers
of $2_{q}$ and $\kappa.$ We ignore them at the moment and will come
back to them later, when they will play a role of corresponding transition
rates of RPM.

\begin{figure}
\centering{}\includegraphics[width=0.4\textwidth]{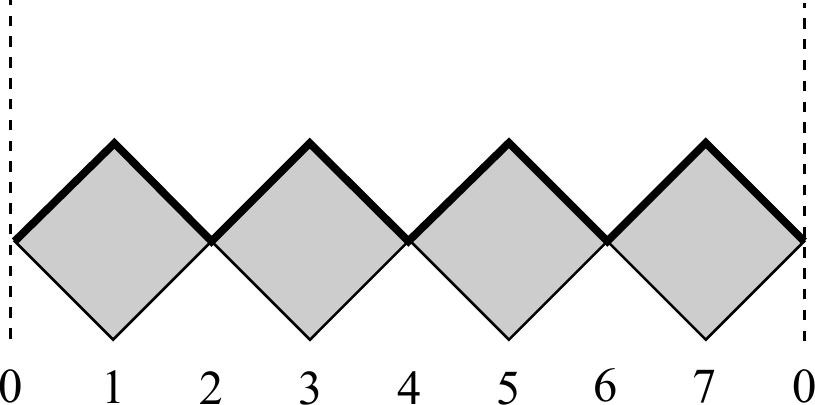}\caption{\label{fig1}The substrate $J_{L}$ of width $L=8$. The cylinder
is cut along vertical line at horizontal position $0$. }
\end{figure}

Before going to the dynamics of the process let us give a pictorial
representation to monomials from $PTL_{L}$. A rhombic tile with a
diagonal of length two is associated with every generator $e_{i}$.
The number $i$ refers to the horizontal position of the center of the
tile in the vertical strip of  width $L$. The periodicity implies
that the strip is wrapped around the cylinder. Then the picture representing
any monomial from the ideal is obtained by adding tiles onto the substrate
$J_{L}$ shown in fig.\ref{fig1} in the same order as the generators
enter into the monomial.

\begin{figure}[b]
\centering{}\includegraphics[width=0.6\paperwidth]{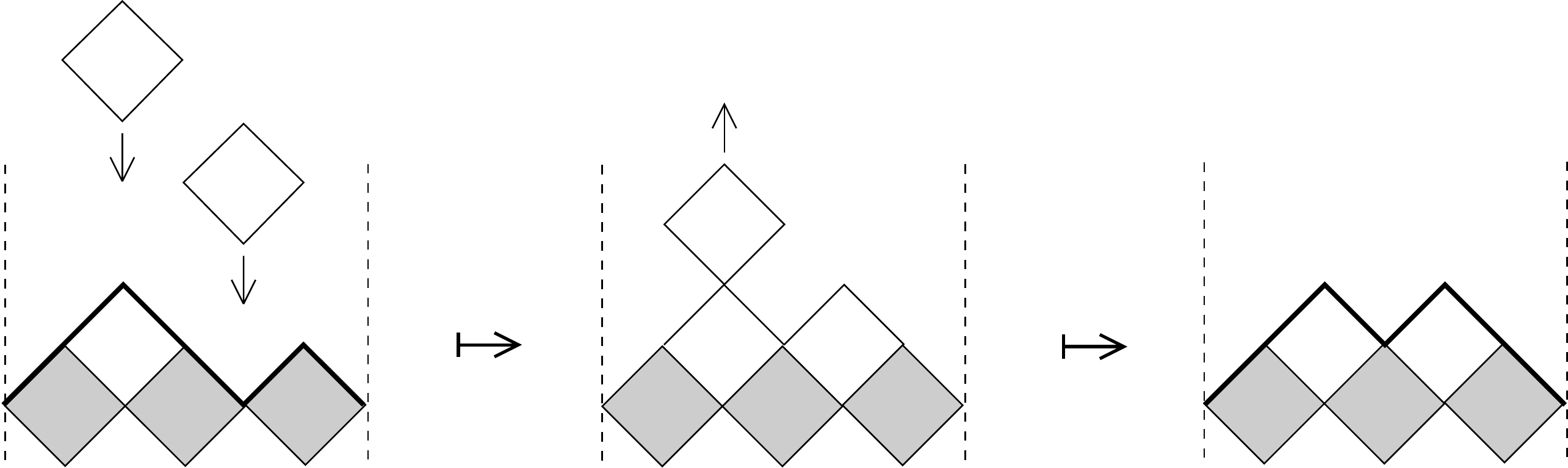}\caption{\label{fig2} Action of two generators $e_{2}$ (left dropping tile)
and $e_{4}$ (right dropping tile) on the state $e_{2}J_{6}$ results
(up to numeric factor $2_{q}$) in configuration $e_{2}e_{4}J_{6}$}
\end{figure}

\begin{figure}
\centering{}\includegraphics[width=0.6\paperwidth]{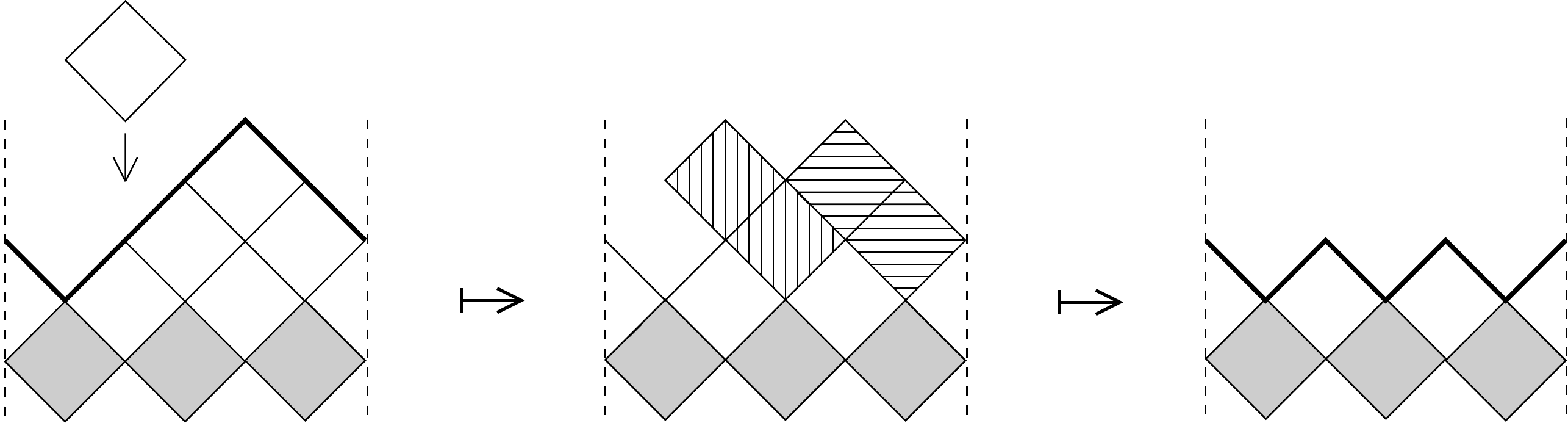}\caption{\label{fig3} Action of generator $e_{2}$ on the state $e_{4}(e_{3}e_{5})(e_{2}e_{4}e_{6})J_{6}$
results in a stable configuration $(e_{2}e_{4}e_{6})J_{6}$. Here
relation (\ref{eq: TL1}) was applied two times causing avalanche
of a size $4$ (we include dropping tile in counting the number of
erased tiles). In general, application of (\ref{eq: TL1}) $k$ times
causes avalanche of a size $2k$. }
\end{figure}

\begin{figure}
\centering{}\includegraphics[width=0.6\paperwidth]{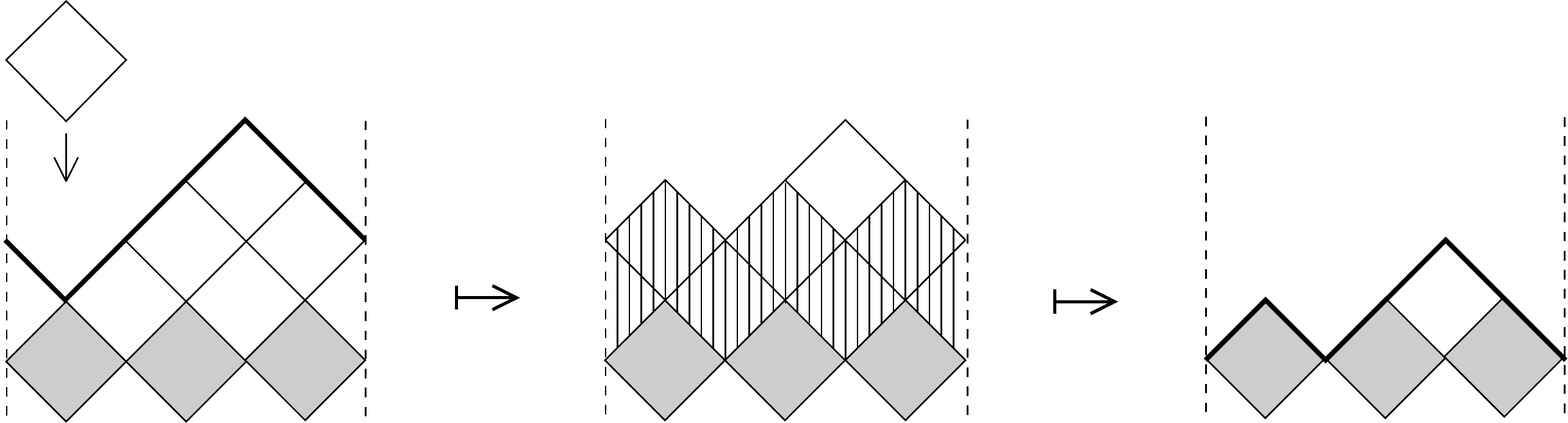}\caption{\label{fig4}Action of generator $e_{1}$ on the state $e_{4}(e_{3}e_{5})(e_{2}e_{4}e_{6})J_{6}$
resulting in a global avalanche in the system of size $L=6$. Algebraically we have $e_{1}\times e_{4}(e_{3}e_{5})(e_{2}e_{4}e_{6})J_{6}=e_4 J_LI_LJ_L=\kappa e_4J_L$, where we use commutativity (\ref{eq: TL3}) on the first step and eq.(\ref{eq: TL_periodic}) on the second.  Pictorially,   the addition of the tile corresponding to  $e_{1}$ completes two full layers of tiles, which are then   removed. The removal of the two layers  is interpreted as the global avalanche of size $L$.}
\end{figure}

The tiles are placed from above to the lowest possible position that
respects the boundaries of the other tiles. Conversely, to reconstruct
a monomial from the picture one has to read the generators off the
tiles going from left to right and from top to bottom (see examples
in figures \ref{fig2}-\ref{fig4}).

An addition of a new tile produces a configuration, which can be either
\textit{unstable} (reducible) or \textit{stable} (irreducible), i.e.
the number of tiles in the system can or can not be reduced by application
of relations (\ref{eq: TL1}-\ref{eq: TL_periodic}) respectively.
Namely, a configuration remains stable, when the tile comes to a downward
corner, the \textit{valley} \textendash{} see action of $e_{4}$ in
fig.2. In contrast, when the tile is placed onto an upward corner,
the \textit{peak}, it should be simply removed according to (\ref{eq: TL1})
\textendash{} see action of $e_{2}$ in fig.2. The latter event is
treated as a {\em reflection} of the tile. If the tile comes neither
to a valley nor to a peak, it is on the \textit{slope} of a ``mountain'',
which suggests, that the relation (\ref{eq: TL2}) is to be applied.
Successive applications of this relation cause an upward \textit{avalanche},
which peels a one tile thick strip off the surface of the ``mountain''
at the level of  arrival of the tile and above \textendash{} see
fig.3. In addition, if the two full layers of tiles have been completed
on top of the substrate, they should be removed according to (\ref{eq: TL_periodic}).
An example of such a global avalanche is shown in fig.4. Note that
all the above transformations preserve  tiles at the level of substrate.
All the changes are limited to the half-strip above the horizontal
mid-line of the substrate, which we refer to as zero level. The upper
boundary of a stable configuration is a closed broken line, touching
the substrate at least once and wrapped around the cylinder. Similar
representation for usual (non-periodic) TL algebra, suggests that
an upper boundary of a stable configurations is a Dyck path that starts
and ends at zero level, staying in the upper half-strip in-between.
Hence, this representation is known as the Dyck path representation.
By analogy we adopt the term \textit{periodic Dyck path} representation
also for the periodic case.

\subsection{Raise and peel model\label{subsec:Raise-and-peel}}

Now we are in a position to formulate the dynamics of RPM in terms of
tiles falling onto the substrate. The continuous time Markov process
starting from the substrate is defined as follows. The system is supplied
with $L$ independent Poissonian clocks at each integer horizontal
position $i=1,\dots,L$, which ring after exponentially distributed
waiting time $\mathbb{P}(t_{i}>\tau)=e^{-\tau}$. When the clock at
position $i$ has rang a tile falls down from above onto a current
configuration at the position $i$. In terms of the algebra the monomial
corresponding to the system configuration is multiplied by $e_{i}$.
The subsequent processes is the reduction of the configuration specified
by relations (\ref{eq: TL1}-\ref{eq: TL_periodic}). Namely, if the
tile comes to a peak it is rejected and disappears. If it comes to
a valley it stays there with a single exception of the tile having
completed two full layers of tiles on top of the substrate. In the
latter case, the two layers are removed. This process affects the
whole system and, hence, is referred to as \textit{global avalanche}.
Also, when the tile comes to the slope, it causes the \textit{local
avalanche}, which starts going up the mountain, peeling off the one-tile-thick
layer, and continues in the same direction up to the moment when it
reaches the same vertical level as it started from. All these processes
are instantaneous in the Poissonian time scale.

For a general continuous time Markov process the stochastic evolution
of configuration $C(t)$ of the system can be characterized by the dependence
of the expectation $\mathbb{E}f(C(t))$ of an arbitrary function $f(C(t))$
on time $t$. This time dependence can be translated to the time dependence
of the function itself
\begin{equation}
\mathbb{E}f(C(t))=\mathbb{E}_{\mathbb{P}_{0}}f_{t}(C),\label{eq: expectation}
\end{equation}
where the expectation in the rhs is with respect to the initial probability
measure $\mathbb{P}_{0}(C):=\mathrm{Prob}(C(t=0)=C)$. The time dependence
of function $f_{t}(C)$ is governed by the backward Kolmogorov equation
\begin{equation}
\partial_{t}f_{t}(C)=\mathcal{L}f_{t}(C),\label{eq:Kolmogorov backward}
\end{equation}
subject to initial conditions $f_{0}(C)=f(C)$, where the action of
backward generator $\mathcal{L}$ is defined by
\begin{equation}
\mathcal{L}f(C)=\sum_{\{C'\}}u_{C,C'}(f(C')-f(C)),\label{eq:backward generator}
\end{equation}
with $u_{C,C'}$ being the transition rates from $C$ to $C'$ and
the sum is over the set of all stable configurations. In particular,
taking the function $f(C')=\delta_{C',C}$ and evaluating the expectation
of (\ref{eq:Kolmogorov backward}) we obtain the forward equation
for the probability $\mathbb{P}_{t}(C)=\mathrm{Prob}(C(t)=C)=\mathbb{E}\delta_{C(t),C}$
\[
\partial_{t}\mathbb{P}_{t}(C)=\mathcal{L}^*\mathbb{P}_{t}(C):=\sum_{\{C'\}}(u_{C',C}\mathbb{P}_{t}(C')-u_{C,C'}\mathbb{P}_{t}(C)).
\]
Conversely, the action of the forward generator $\mathcal{L}^*$ on
$\delta$-measures $\delta_{C}(C')=\delta_{C,C'}$, which form the
natural basis of the space of linear functionals on functions $f(C)$,
is
\begin{equation}
\mathcal{L}^*\delta_{C}=\sum_{\{C'\}}u_{C,C'}(\delta_{C'}-\delta_{C}).\label{eq:L delta}
\end{equation}

In the case of RPM the space of linear functionals can be naturally
identified with $\mathcal{J}_{L}$. Then the linear operator $\mathcal{L}^*$
can be realized within the ideal $\mathcal{J}_{L}$ as an operator
of multiplication by an element
\begin{equation}
\mathcal{L}^*=\sum_{i=1}^{L}(e_{i}-1).\label{RPM-H}
\end{equation}
Indeed, for every monomial $m_{C}$ representing a stable configuration
$C$ we have $\mathcal{L}^*m_{C}=\sum_{\{C'\}}(m_{C'}-m_{C})$, where
$m_{C'}=e_{i}m_{C}$ for $i=1,\dots,L$. This exactly coincides with
(\ref{eq:L delta}) given the nonzero rates are $u_{C,C'}=1$ and
the reduction of $e_{i}m_{C}$ to its stable form yields no any numeric
coefficients. As it was mentioned above, an application of the relations
(\ref{eq: TL1}-\ref{eq: TL_periodic}) yields integer powers of parameters
$\kappa$ and $2_{q}$. Therefore to maintain the stochasticity we
have to choose the parameters to be
\begin{equation}
\kappa=1,\,\,\,q=e^{i\pi/3}.\label{eq:stochastic}
\end{equation}
\begin{rem}
{\small{}{}As was discussed, the functions on configurations and
the probability distributions, aka linear functionals on them, are
dual to each other. The Markov process can be described in terms of
time-dependence of either of them, which in turn is defined by backward
or forward generators respectively. The readers used to the language
of quantum mechanics can define the ``state vector''
\[
\left|P_{t}\right\rangle =\sum_{\{C\}}\mathbb{P}_{t}(C)m_{C}=\sum_{\{C\}}\mathbb{P}_{0}(C)m_{C}(t)
\]
from $\mathcal{J}_{L},$ whose evolution can be attributed either
to time dependence of coordinates or that of the basis vectors defined
by $\mathcal{L}$ or $\mathcal{L}^*$, respectively. }
\end{rem}

To conclude the formulation of RPM we note that there are two other
pictorial representations of the TL algebra in terms of links and
in terms of particles, which can be useful for visualizing the algebraic
relations and the dynamics of the model.
For the link presentation we refer the reader
to \citep{AR-07}. The particle representation will be briefly discussed later in Section \ref{sec: discussion} within the discussion of  our results.

\subsection{R-matrix representation}

The algebraic construction we have been discussing so far is not very
useful for analytic calculations. For practical purposes it is more
convenient to work with $R$-matrix representation of PTL algebra.
We consider the representation of $PTL_{L}$ on the space $\mathcal{H}=V^{\otimes L}$
obtained by tensoring $L$ copies of $V=\mathbb{C}^{2}.$ Specifically,
we use the matrix
\[
\check{R}:=\left(\begin{array}{cccc}
q & 0 & 0 & 0\\
0 & 0 & u^{-1} & 0\\
0 & u & q-q^{-1} & 0\\
0 & 0 & 0 & q
\end{array}\right)
\]
depending on the deformation and twist parameters $q,u\in\mathbb{C}$,
which obeys a quadratic relation known as the {\em Hecke condition}
\begin{equation}
\check{R}^{2}=\boldsymbol{1}+(q-q^{-1})\check{R}.\label{eq:Hecke}
\end{equation}
The matrix can be thought of as a representation of an operator acting
in the tensor product of two two-dimensional spaces $V\otimes V$.
To construct a generalization acting on the whole space $\mathcal{H}$
we introduce operator
\[
\check{R}_{i,i+1}=\boldsymbol{1}_{1}\otimes\cdots\otimes\boldsymbol{1}_{i-1}\otimes\check{R}\otimes\mathbf{1}_{i+2}\otimes\cdots\otimes\boldsymbol{1}_{L}
\]
that acts as an identity operator on all the components of the tensor
product except the spaces with numbers $i,i+1\,\,\,\mathrm{mod}\,\,\,L$
intertwined by $\check{R}.$ In addition to the Hecke relation these
operators also satisfy the \textit{ braid relation}\footnote{Multiplying $\check{R}_{i,j}$ by permutation matrix $P_{i,j}$ that
interchanges spaces $i,j$ one obtains another R-matrix $R_{i,j}=P_{i,j}\check{R}_{i,j}$,
which is of more common use in context of quantum groups for satisfying
the\textit{ Yang-Baxter relation} $R_{1,2}R_{1,3}R_{2,3}=R_{2,3}R_{1,3}R_{1,2}.$}
\begin{equation}
\check{R}_{i,i+1}\check{R}_{i+1,i+2}\check{R}_{i,i+1}=\check{R}_{i+1,i+2}\check{R}_{i,i+1}\check{R}_{i+1,i+2},\label{eq:Braid}
\end{equation}
where all the sub-indices should be read $\mathrm{mod}\,\,L$, while
operators $\check{R}_{i,i+1}$and $\check{R}_{j,j+1}$ acting non-trivially
on spaces, which are not next to each other, commute
\begin{equation}
\check{R}_{i,i+1}\check{R}_{j,j+1}=\check{R}_{j,j+1}\check{R}_{i,i+1},\,\,\,|i-j|>1\,\,\mbox{(mod \ensuremath{L}).}\label{eq:R-commute}
\end{equation}
To define the representation $\rho:PTL_{L}\to\mathrm{End}\,\mathcal{H}$
as an algebra homomorphism from $PTL_{L}$ to linear transformations
of $\mathcal{H},$ we assign
\[
\rho(e_{i})=q\mathbf{1}-\check{R}_{i,i+1}.
\]
With this prescription it is not difficult to check that relations
(\ref{eq:Hecke}-\ref{eq:R-commute}) are indeed equivalent to relations
(\ref{eq: TL1}-\ref{eq: TL3}) between generators of $PTL_{L}$ respectively.
A bit more involved calculations show that relation (\ref{eq: TL_periodic})
also holds with
\begin{equation}
\kappa=\left(u^{L/2}+u^{-L/2}\right)^{2}\label{eq:alpha vs u}
\end{equation}
We conclude this subsection by writing down the form that the forward
Markov generator (\ref{RPM-H}) takes in the $R$-matrix representation

\[
\rho(\mathcal{L}^*)=\sum_{i=1}^{L}\left[(q-1)\mathbf{1}-\check{R}_{i,i+1}\right].
\]
Despite it represents the stochastic operator, the basis this matrix
is written in is the spin basis rather than the basis of RPM configurations.
This is why the matrix $\rho(\mathcal{L}^*)$ is not stochastic even
when the stochasticity condition (\ref{eq:stochastic}) is satisfied.

\subsection{XXZ model}

The $R$-matrix form of the forward generator takes the familiar form
of the Hamiltonian of quantum XXZ chain if we rewrite the $R$-matrices
in terms of $\sigma-$matrices
\[
\sigma^{+}=\left(\begin{array}{cc}
0 & 1\\
0 & 0
\end{array}\right),\,\,\,\sigma^{-}=\left(\begin{array}{cc}
0 & 0\\
1 & 0
\end{array}\right),\,\,\,\sigma^{z}=\left(\begin{array}{cc}
1 & 0\\
0 & -1
\end{array}\right).
\]
The matrix representing generator $e_{i}$ can be expanded into a
sum of tensor products of such matrices acting non-trivially on the
pair of spaces $V_{i}\otimes V_{i+1}$ from $\mathcal{H}$

\[
\hspace{-2.5cm}\rho(e_{i})=\left[u\sigma_{i}^{+}\otimes\sigma_{i+1}^{-}+u^{-1}\sigma_{i}^{-}\otimes\sigma_{i+1}^{+}-\frac{\cos\gamma}{2}(\sigma_{i}^{z}\otimes\sigma_{i+1}^{z}-\mathbf{1}\otimes\mathbf{1})+\frac{\mathrm{i}\sin\gamma}{2}(\sigma_{i}^{z}\otimes\mathbf{1}-\boldsymbol{1}\otimes\sigma_{i+1}^{z})\right],
\]
where the sub-indices of $\sigma-$matrices index the tensor components
of $\mathcal{H}$, where these matrices act, and we introduce new notation
$\gamma$ by
\begin{equation}
\cos\gamma=\frac{2_{q}}{2},\qquad\sin\gamma=\frac{q-q^{-1}}{2\mathrm{i}}.\label{eq:gamma vs q}
\end{equation}
In this form it is easy to see that the forward generator of Markov process
is represented by
\begin{equation}
\rho(\mathcal{L}^*)=-\left(2-\cos\gamma\right)L/2-H_{XXZ}^{\Delta,u},\label{eq:L=00003D00003DH_XXZ}
\end{equation}
where
\begin{equation}
H_{XXZ}^{\Delta,u}=-\sum_{i=1}^{L}\left[u\sigma_{i}^{+}\otimes\sigma_{i+1}^{-}+u^{-1}\sigma_{i}^{-}\otimes\sigma_{i+1}^{+}+\frac{\Delta}{2}\sigma_{i}^{z}\otimes\sigma_{i+1}^{z}\right]\label{H-XXZ}
\end{equation}
is the Hamiltonian of the antiferromagnetic Heisenberg quantum chain
of $L$ spins 1/2 with anisotropy
\begin{equation}
\Delta=-\cos\gamma\label{eq:Delta vs gamma}
\end{equation}
and the twist parameter $u$, which accounts for the effect of magnetic
flux through the ring when $|u|=1$. In case $\left|u\right|\neq1$,
the latter breaks the left-right symmetry of the model making the
Hamiltonian non-Hermitian. Here we imply the periodic boundary conditions
$\left\{ \sigma_{L+1}^{\pm}=\sigma_{1}^{\pm},\,\sigma_{L+1}^{z}=\sigma_{1}^{z}\right\} $.
Alternatively, by a unitary transformation the parameter $u$ can
be transferred from the Hamiltonian to twisted boundary conditions
\citep{ABB}:
\[
\left\{ \sigma_{L+1}^{\pm}=u^{\pm L}\sigma_{1}^{\pm},\,\,\,\sigma_{L+1}^{z}=\sigma_{1}^{z}\right\} .
\]

To describe the relation between RPM and XXZ model precisely we should
specify the subspace of $\mathcal{H}$ that is isomorphic to $\mathcal{J}_{L}$.
As is well known \citep{ABB} the XXZ Hamiltonian (\ref{H-XXZ}) commutes
with an operator of the total spin projection $S_{z}=\frac{1}{2}\sum_{i=1}^{L}\sigma_{i}^{z}$.
This is why the invariant subspaces of $H_{XXZ}^{\Delta,u}$ are indexed
by values of the total projection $S_{z}\in\{-L/2,\dots L/2\}$. In particular,
the periodic Dyck path representation is isomorphic to the subspace
with $S_{z}=0$.

\subsection{Bethe ansatz }

In the following we will be interested in eigenvalues of the generators
of RPM $\mathcal{L}^*$ and $\mathcal{L}$, which are related to those from the $S_{z}=0$
sector of $H_{XXZ}^{\Delta,u}$ due to (\ref{eq:L=00003D00003DH_XXZ}).

As is well known, the XXZ Hamiltonian is diagonalized in each sector
of a given total spin projection $S_{z}=L/2-m$, $m=0,1,\dots,L$,
by the Bethe eigenvectors
\[
\left|\boldsymbol{z}\right\rangle =\sum_{1\leq i_{1}<\dots<i_{m}\leq L}\psi(i_{1},\dots,i_{m}|\boldsymbol{z})\sigma_{i_{1}}^{-}\otimes\dots\otimes\sigma_{i_{m}}^{-}\left|\left(\begin{array}{c}
1\\
0
\end{array}\right)\otimes\dots\otimes\left(\begin{array}{c}
1\\
0
\end{array}\right)\right\rangle ,
\]
where the function
\[
\psi(i_{1},\dots,i_{m}|\boldsymbol{z})=\sum_{\sigma\in S_{m}}\left(-1\right)^{\sigma}\prod_{1\leq i<j\leq m}\frac{1-2u\Delta z_{\sigma(i)}+u^{2}z_{\sigma(i)}z_{\sigma(j)}}{1-2u\Delta z_{i}+u^{2}z_{i}z_{j}}\prod_{k=1}^{m}z_{\sigma(k)}^{i_{k}}
\]
is a sum over permutations $\sigma\in S_{m}$ of $m!$ terms parametrized
by a particular solution $\boldsymbol{z}=(z_{1},\dots,z_{m})$ of
a system of the Bethe equations

\[
z_{i}^{L}=\left(-1\right)^{m-1}\prod_{j=1}^{m}\frac{1-2u\Delta z_{i}+u^{2}z_{i}z_{j}}{1-2u\Delta z_{j}+u^{2}z_{i}z_{j}},\,\,\,i=1,\dots,m.
\]
Every solution defines a single eigenvector corresponding to the eigenvalue
of the Hamiltonian, i.e. the energy

\[
E_{XXZ}=\Delta\left(2m-\frac{L}{2}\right)-\sum_{i=1}^{m}\left[uz_{i}+\left(uz_{i}\right)^{-1}\right].
\]

In general evaluation of a particular value of the energy is a difficult
problem that requires finding a particular solution of the system
explicitly. However, the relation
\[
\Lambda^{RPM}=-E_{XXZ}-\left(2+\Delta\right)L/2,
\]
between eigenvalue $\Lambda^{RPM}$ of the stochastic generators $\mathcal{L}$
and $\mathcal{L}^{*}$ and the energy $E_{XXZ}$ that follows from
(\ref{eq:L=00003D00003DH_XXZ}) allows one to get some immediate information
relying on the stochasticity only. Specifically a generator of Markov
process has a non-degenerate zero eigenvalue
\[
\Lambda_{0}^{RPM}=0.
\]
which is the eigenvalue with the largest real part. Thus, the minimal
energy of the XXZ model in the sector $S_{z}=0$ at the stochastic
point (\ref{eq:stochastic}), i.e
\begin{equation}
\gamma=\pi/3,\quad\Delta=-1/2,\quad u=e^{\mathrm{2\pi i}/(3L)},\label{eq: stochastic gamma, Delta, u}
\end{equation}
is $E_{XXZ}=-3L/4$, which is also the total minimum over all sectors.
Remarkably, though the most of results on the energy levels of XXZ
model are obtained in the thermodynamic limit $L\to\infty$, the latter
one is exact for arbitrary $L$. In the following we will go beyond
the stochastic point, using this value as the starting point of the
large deviation analysis.

\subsection{Stationary state}

It was first observed by Razumov and Stroganov \citep{RS-01} that
the ground state eigenvector of the XXZ Hamiltonian at $\Delta=-1/2$
possesses  remarkable combinatorial properties. This observation
was immediately reinterpreted and further developed in applying
to the ground state vector of the dense $O(1)$ model \citep{BGN,RS-04,RS-05}
which, in turn, led to a formulation of the stochastic Raise and Peel
model \citep{GNPR}. An extensive list of conjectures relating stationary
probabilities of the RPM with enumerations of various symmetry classes
of the Alternating Sign Matrices and Plane Partitions was later presented
in \citep{MNGB,P}. Below we formulate a pair of such conjectures.
Their large scale limits will be approved later in section 3.2 (see
eqs.(\ref{eq: c1 tile}),(\ref{approve2})).

\begin{figure}
\centering{}\includegraphics[width=0.65\paperwidth]{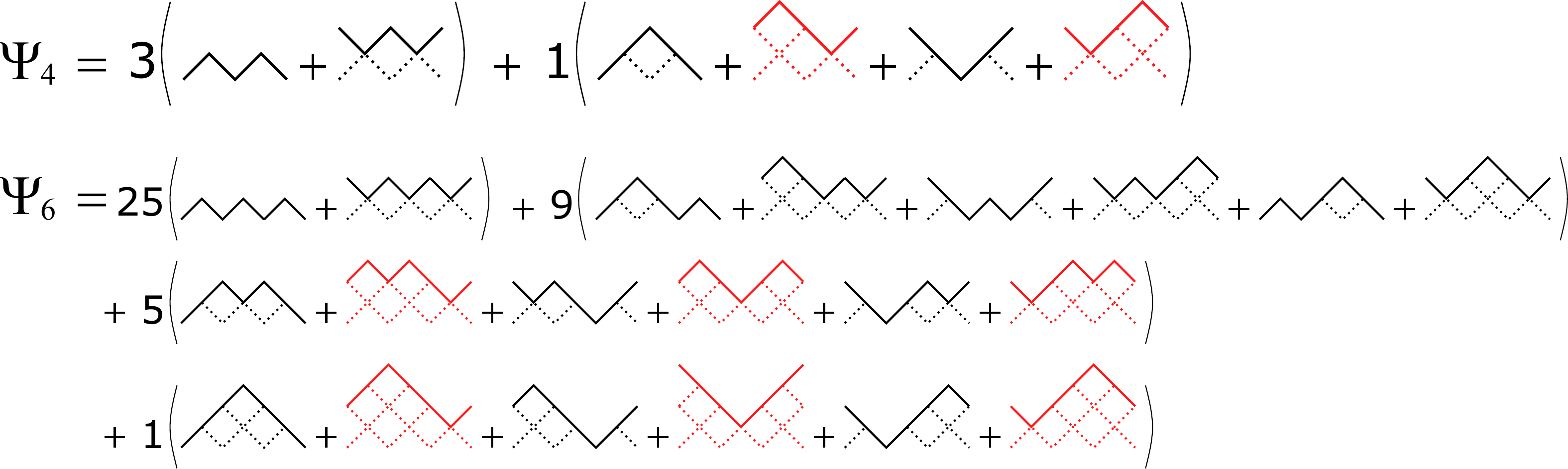}\caption{\label{fig5} Stationary states of the Raise and Peel Model for $L=4$
and $L=6$. }
\end{figure}

Denote $\Psi_{L}$ a vector in the ideal $\mathcal{J}_{L}$ which
corresponds to the stationary state of the RPM: $\mathcal{L}^*\Psi_{L}=0$.
Written in the basis of the periodic Dyck paths this vector gives
stationary probabilities of the stable configurations. Combinatorial
properties of the stationary state are most suitably observed from
$\Psi_{L}$ if it is normalized so that its smallest coefficients
are equal to 1. In this case a total sum of all coefficients of $\Psi_{L}$,
i.e. the probability normalization, coincides with the number of
$L\times L$ half turn symmetric Alternating Sign Matrices \citep{MNGB}:
\[
A_{\mbox{\tiny HT}}(L)=\prod_{k=0}^{L/2-1}\frac{(3k+2)!(3k)!}{(L/2+k)!^{2}}=2,10,140,\dots
\]
In figure \ref{fig5} we present explicit expressions for $\Psi_{L}$
for small values of $L$. One can use them to test the formula above
and the conjectures given below.

\textbf{Conjecture 1.}\citep{MNGB} An average number of peaks (local
maxima) as well as valleys (local minima) in the stationary state
$\Psi_{L}$ equals
\[
L\cdot\frac{3L^{2}}{8(L^{2}-1)}\,=\,2\cdot\mbox{\ensuremath{\frac{1}{2}}},\,\,4\cdot\mbox{\ensuremath{\frac{2}{5}}},\,\,6\cdot\mbox{\ensuremath{\frac{27}{70}}},\dots\quad\mbox{for \ensuremath{L=2,4,6,\dots}}
\]

\textbf{Conjecture 2.} Stable configurations which do not have valleys
on level $0$ and have a single valley on level $1$ (these configurations
are painted red in fig.\ref{fig5}) appear in the stationary state
$\Psi_{L}$ with probability
\[
\frac{3L}{4(L^{2}-1)}\,=\,\frac{1}{2},\,\,\frac{1}{5},\,\,\frac{9}{70},\dots\quad\mbox{for \ensuremath{L=2,4,6,\dots}}
\]

From the above formulas one infers the expressions for the total number
of tiles removed within the avalanches and the number of global avalanches
per unit time in the large time limit:
\begin{equation}
\lim_{t\to\infty}\mathbb{E}\frac{\mathcal{N}_{t}^{\lozenge}}{t}=L\cdot\frac{5L^{2}-8}{8(L^{2}-1)},\label{eq: mean total}
\end{equation}
\begin{equation}
\lim_{t\to\infty}\mathbb{E}\frac{\mathcal{N}_{t}^{\circlearrowleft}}{t}=\frac{3}{4}\frac{L}{L^{2}-1}.\label{eq: mean global}
\end{equation}
Here for the first formula one notices that in the stationary state
the avalanches should remove all the tiles that are not reflected,
i.e. those hitting configurations everywhere except the peaks. For
the second formula one observes that the states described in Conjecture
2 are precisely those which may suffer the global avalanches.

\section{Large deviations}

In subsection \ref{subsec:Raise-and-peel} we discussed evaluation
of expectations of functions on configurations of Markov process at
fixed moment of time. They are given in terms of the solution of the
backward equation (\ref{eq: expectation},\ref{eq:Kolmogorov backward}),
which in turn can be represented as the integral over trajectories
of the process

\begin{eqnarray}
\mathbb{E}f(C(t)) & = & \mathbb{E}_{\mathbb{P}_{0}}e^{t\mathcal{L}}f(C)\label{eq:expect vs generator}\\
 & = & \sum_{n,C_{0},\dots,C_{n}}f(C_{n})\int_{0\leq t_{1}\leq\dots\leq t_{n}\leq t}d\mathbb{P}_{C_{0},,\dots,C_{n}}(t_{1},\dots,t_{n}),\label{eq:integral over trajectories}
\end{eqnarray}
where

\[
d\mathbb{P}_{C_{0},\dots,C_{n}}(t_{1},\dots,t_{n})=\mathbb{P}_{0}(C_{0})\prod_{k=0}^{n-1}u_{C_{k},C_{k+1}}e^{-\left(t_{k+1}-t_{k}\right)u_{C_{k},C_{k+1}}}dt_{k+1}
\]
is the measure of the trajectories starting from initial distribution
$\mathrm{Prob}(C(t=0)=C)=\mathbb{P}_{0}(C)$ at time $t_{0}=0$, passing
through configurations $C_{0},\dots,C_{n}$ by time $t$ so that the
jump from $C_{i-1}$to $C_{i}$ has happened within interval $[t_{i},t_{i}+dt_{i}]$
for $i=1,\dots,n$.

In similar fashion one can study the statistics of additive functionals
on the system trajectories. Specifically, consider a right-continuous
random (possibly multicomponent) function of time $\boldsymbol{\mathcal{N}}_{t}=(\mathcal{N}_{t;1},\dots,\mathcal{N}_{t;k})$,
which initially is $\boldsymbol{\mathcal{N}}_{t=0}=(0,\dots,0)$ and
changes abruptly by
\[
\boldsymbol{\mathcal{N}}_{t}=\boldsymbol{\mathcal{N}}_{t-0}+\delta\boldsymbol{\mathcal{N}}_{C,C'}
\]
every time the system jumps between two configurations. The values
of $\delta\boldsymbol{\mathcal{N}}{}_{C,C'}$, fixed once for all,
depend only on configurations $C$ and $C'$ before and after the
jump respectively, but not on time or history of the process. Hence,
for the trajectory that has passed through configurations $C_{0},\dots,C_{n}$
by time $t$, we have $\boldsymbol{\mathcal{N}}_{t}=\delta\boldsymbol{\mathcal{N}}_{C_{0},C_{1}}+\cdots+\delta\boldsymbol{\mathcal{N}}_{C_{n-1},C_{n}}$.
Therefore, the expectations of the form $\mathbb{E}\left(f(C_{t})e^{\boldsymbol{\alpha}\boldsymbol{\mathcal{N}}_{t}}\right)$,
where $\boldsymbol{\alpha}=(\alpha_{1},\dots,\alpha_{k})$ is a parameter
and $\boldsymbol{\alpha}\boldsymbol{\mathcal{N}}_{t}=\sum_{i}\alpha_{i}\mathcal{N}_{t;i}$,
can be written in the form of integral over trajectories, similar
to (\ref{eq:integral over trajectories}) up to the change of integration
measure $d\mathbb{P}_{C_{0},,\dots,C_{n}}(t_{1},\dots,t_{n})$ to
a new measure
\[
\hspace{-12mm}d\mathbb{M}_{C_{0},\dots,C_{n}}^{\boldsymbol{\alpha}}(t_{1},\dots,t_{n})=\mathbb{P}_{0}(C_{0})\prod_{k=0}^{n-1}\left[u_{C_{k},C_{k+1}}e^{\boldsymbol{\alpha}\delta\boldsymbol{\mathcal{N}}_{C_{k},C_{k+1}}}\right]e^{-\left(t_{k+1}-t_{k}\right)u_{C_{k},C_{k+1}}}dt_{k+1}
\]
obtained from the former one by assigning an additional weight $e^{\boldsymbol{\alpha}\delta\mathcal{\boldsymbol{N}}_{C_{k},C_{k+1}}}$
to every jump between two configurations. Of course, this is not a
probability measure anymore, i.e. it does not have a unit normalization.
Its total normalization obtained by setting $f(C)\equiv1$ is exactly
the quantity of our interest, the generating function $\mathbb{E}e^{\boldsymbol{\alpha}\boldsymbol{\mathcal{N}}_{t}}$
of joint moments of $\boldsymbol{\mathcal{N}}_{t}$. Similarly to
(\ref{eq:expect vs generator}) it can be written in terms of the
exponential
\begin{equation}
\mathbb{E}e^{\boldsymbol{\alpha}\boldsymbol{\mathcal{N}}_{t}}=\mathbb{E}_{\mathbb{P}_{0}}\left(e^{t\mathcal{L}_{\boldsymbol{\alpha}}}\boldsymbol{1}\right),\label{eq: gen function}
\end{equation}
of a new operator $\mathcal{L}_{\alpha}$,
\[
\mathcal{L}_{\boldsymbol{\alpha}}f(C)=\sum_{C'\neq C}u_{C,C'}\left(e^{\boldsymbol{\alpha}\delta\boldsymbol{\mathcal{N}}_{C,C'}}f(C')-f(C)\right).
\]
The main point of the large deviation theory applied to the finite
state space Markov process is the observation that in the large time
limit the exponential in the r.h.s. of (\ref{eq: gen function}) is dominated
by the largest eigenvalue of the generator $\mathcal{L}_{\boldsymbol{\alpha}}$,
that is to say that the scaled CGF converges to the largest eigenvalue
\[
I_{\boldsymbol{\mathcal{N}}_{t}}(\boldsymbol{\alpha}):=\lim_{t\to\infty}t^{-1}\mathbb{\log E}e^{\boldsymbol{\alpha}\boldsymbol{\mathcal{N}}_{t}}=\Lambda_{0}(\boldsymbol{\alpha}).
\]

Returning back to RPM we are going to analyze large deviations of
two-component additive functional $\boldsymbol{\mathcal{N}}_{t}=\left(\mathcal{N}_{t}^{\circlearrowleft},\mathcal{N}_{t}^{\lozenge}\right)$
on the trajectories of the process, where
\begin{description}
\item [{$\mathcal{N}_{t}^{\circlearrowleft}$-}] the total number of global
avalanches occurred in the system by the time $t$.
\item [{$\mathcal{N}_{t}^{\lozenge}$-}] the total number of tiles removed
from the systems during avalanches (both local and global) by the
time $t$,
\end{description}
Following the above discussion, we would like to obtain the scaled
joint CGF of $\left(N_{t}^{\circlearrowleft},\mathcal{N}_{t}^{\lozenge}\right)$,
i.e. the largest eigenvalue $\Lambda(\alpha,\beta)$ of the operator
$\mathcal{L}_{(\alpha,\beta)}$ obtained from the generator (\ref{eq:backward generator})
by multiplying the off-diagonal matrix elements $u_{C,C'}$ by $e^{\alpha\delta\mathcal{N}_{C,C'}^{\circlearrowleft}+\beta\delta\mathcal{N}_{C,C'}^{\lozenge}}$.

To this end, we extend the algebraic construction of the previous
section to the operator $\mathcal{L}_{(\alpha,\beta)}$. First, let
us rescale generators of the periodic Temperley-Lieb algebra
\[
\tilde{e}_{i}:=2_{q}^{-1}e_{i},\quad\tilde{I}_{L}:=\tilde{e}_{2}\tilde{e}_{4}\dots\tilde{e}_{L},\quad\tilde{J}_{L}:=\tilde{e}_{1}\tilde{e}_{3}\dots\tilde{e}_{L-1}.
\]
PTL relations (\ref{eq: TL1}-\ref{eq: TL_periodic}) in terms of
them read:
\begin{eqnarray}
\tilde{e}_{i}^{2} & = & \tilde{e}_{i},\label{eq: TL1-1}\\
\tilde{e}_{i}\tilde{e}_{i\pm1}\tilde{e}_{i} & = & 2_{q}^{-2}\tilde{e}_{i},\label{eq: TL2-1}\\
\tilde{e}_{i}\tilde{e}_{j} & = & \tilde{e}_{j}\tilde{e}_{i},\,\,\,\left|i-j\right|>1,\label{eq: TL3-1}\\
\tilde{J}_{L}\tilde{I}_{L}\tilde{J}_{L} & = & \kappa2_{q}^{-L}\tilde{J}_{L},\quad\tilde{I}_{L}\tilde{J}_{L}\tilde{I}_{L}\,\,=\,\,\kappa2_{q}^{-L}\tilde{I}_{L}.\label{eq: TL_periodic-1}
\end{eqnarray}
Note that in this presentation nontrivial numerical factors depending
on two parameters $q$ and $\kappa$ appear in relations (\ref{eq: TL2-1})
and (\ref{eq: TL_periodic-1}) only. These two relations are just
the ones responsible for removing tiles during avalanches: one-time
application of (\ref{eq: TL2-1}) results in removing a pair of tiles
from a configuration within a local avalanche and multiplies corresponding
monomial by $2_{q}^{-2}$ ; (\ref{eq: TL_periodic-1}) removes $L$
tiles within a single global avalanche and multiplies the monomial
by $\kappa2_{q}^{-L}$. The numeric factors enter the matrix elements
of linear operators corresponding to multiplication of monomials in
the ideal $\mathcal{J}_{L}$ by operators $\tilde{e}_{i}.$ In particular,
if we set
\[
\kappa=e^{\alpha}\,\,\,\mathrm{and\,\,\,}2_{q}^{-1}=e^{\beta},
\]
the off-diagonal matrix elements of an operator

\begin{equation}
\mathcal{L}^*_{(\alpha,\beta)}=\sum_{i=1}^{L}(\tilde{e}_{i}-1),\label{eq:L_a,b in I}
\end{equation}
will be those of the forward generator $\mathcal{L}^*$ of RPM multiplied
by additional factors $e^{\alpha\delta\mathcal{N}_{C,C'}^{\circlearrowleft}+\beta\delta\mathcal{N}_{C,C'}^{\lozenge}}$,
as was required. The representation of (\ref{eq:L_a,b in I}) in $\mathcal{H}$
is still given in terms of XXZ Hamiltonian
\[
\rho\left(\mathcal{L}^*_{(\alpha,\beta)}\right)=-e^{\beta}H_{XXZ}^{\Delta,u}-\frac{3L}{4},
\]
where the values of parameters $\Delta$ and $u$ are now away from
the stochastic point (\ref{eq: stochastic gamma, Delta, u}) being
related to $\alpha$ and $\beta$ by
\[
u^{L/2}+u^{-L/2}=e^{\alpha/2},\,\,\,2\Delta=-e^{-\beta}.
\]

With this parametrization we find the largest eigenvalue of the operator
$\mathcal{L}^*_{(\alpha,\beta)}$, aka scaled generating function of
the joint cumulants of $\mathcal{N}_{t}^{\circlearrowleft}$ and $\mathcal{N}_{t}^{\lozenge}$.
\begin{eqnarray}
\Lambda_{0}(\alpha,\beta) & = & -e^{\beta}E_{L}^{XXZ}(\Delta,u)-\frac{3L}{4}.\label{eq: Lambda_0(alpha,beta)}
\end{eqnarray}

\subsection{The groundstate of XXZ Hamiltonian\label{subsec:The-groundstate-of XXZ}}

Let us summarize what is known about the largest eigenvalue of twisted
XXZ Hamiltonian. Here we concentrate on the thermodynamic limit $L\to\infty$
and on the values of the other parameters $\Delta$ and $u$ chosen
such that the related parameters $\alpha$ and $\beta$ are finite,
$\alpha,\beta\in\mathbb{R}$. This suggests that $\Delta\leq0$ and
$u=\exp\left[\mathrm{i}\varphi/L\right]$ with $\varphi$ either real,
$0\leq\varphi\leq\pi$, or pure imaginary, $0\leq\mathrm{i}\varphi<\infty.$
The stochastic point corresponds to $\Delta=-1/2$ and $\varphi=2\pi/3$.

In the thermodynamic limit $E_{L}^{XXZ}(\Delta,u)$ consists of the
bulk term, growing linearly in $L$ as $L\to\infty$ and finite size
corrections (FSC). As anticipated from the scaling of $u$, the bulk
energy per site
\[
e_{\infty}(\Delta)=\lim_{L\to\infty}L^{-1}E_{L}^{XXZ}(\Delta,u),
\]
depends only on $\Delta$, while $\varphi$ appears in the FSC. Their
explicit expressions and methods of derivation depend on which of
two regimes, gapless $|\Delta|<1$ or gapped $\Delta<-1$, is considered.

The bulk term was obtained by Yang and Yang \citep{YY1,YY2} in the
whole range of values of $\Delta$. The FSC where analyzed in a number
of papers in different contexts. Initially, a systematic method of
calculation of FSC was proposed in \citep{dVW} for gapped regime
and extended to gapless regime in \citep{H1}.

The gapless regime attracted a lot of attention in connection with
conformal field theory (CFT) predictions to the free energy of 2D
statistical systems. The FSC to the free energy of models like Potts
model, Ashkin-Teller model, six-vertex and eight-vertex models were
obtained and tested against CFT predictions (see the review \citep{BB}
and references therein). These studies exploited the analogy with
XXZ model at particular fixed values of twist parameter. The general
formula for the groundstate energy as a function of the twist was first
proposed on the basis of numerical solution of Bethe equations in
\citep{ABB}. It was then derived analytically from their finite size
analysis in \citep{HQB}. The CFT meaning of these results obtained
for arbitrary twist was clarified in \citep{DdV}.

The gapped regime is less studied. In addition to \citep{dVW} the
XXZ chain with antiperiodic boundary conditions was considered \cite{H2}, which
corresponds to a particular value of the twist $\varphi=\pi$.

Bellow we give a brief survey of the behaviour of $e_{\infty}(\Delta)$
and of FSC for $\Delta$ and $\varphi$ in the range of interest.
For the gapless phase we refer to a straightforward generalization
of results of \citep{dVW,H1,HQB}. Note that the imaginary twist ceases
the Hamiltonian from being Hermitian. The Bethe roots deviate from
the real axis in this case. This deviation however is itself of order
of $1/L$, assuring that the contour of roots does not encounter any
singularities and all the arguments using contour integration are
preserved.

In the gapped phase no any formulas for arbitrary twist were yet obtained
to our knowledge. Below we give the result obtained by direct extension
of techniques of \citep{dVW,VW}. The detailed calculations will be
presented elsewhere.

We also briefly mention the intermediate scaling regime that connects
the two phases, though no finial exact results were obtained for it
yet.

\subsubsection*{Gapless phase }

In the  gapless regime with
\[
\Delta=-\cos\gamma,\,\,\gamma\in[0,\pi/2]
\]
the bulk part is
\begin{equation}
e_{\infty}(\Delta)=\frac{\cos\gamma}{2}-\sin^{2}\gamma\,Y(\gamma)\label{eq:e_bulk gapless}
\end{equation}
given in terms of an integral
\begin{equation}
Y(\gamma)=\int_{-\infty}^{+\infty}\frac{dx}{\cosh\left(\pi x\right)}\frac{1}{\left[\cosh(2\gamma x)-\cos\gamma\right]}.\label{eq: Y}
\end{equation}
The $\varphi-$dependent FSC of order of $1/L$ are quadratic in $\varphi$
\begin{equation}
\lim_{L\to\infty}L\left(E_{L}^{XXZ}(\Delta,e^{\mathrm{i}\varphi/L})-Le_{\infty}(\Delta)\right)=-\frac{\pi^{2}\sin\gamma}{6\gamma}+\varphi^{2}\frac{\pi\sin\gamma}{4\gamma(\pi-\gamma)},\label{eq:FSC crit}
\end{equation}
where the union of the real and imaginary domains of $\varphi\in[0,\pi]\bigcup\mathrm{i\times}[0,\infty]$
suggests that $-\infty<\varphi^{2}\leq\pi^{2}.$

\subsubsection*{Gapped phase }

In the gapped phase, $\Delta<-1$ we use the parametrization
\[
\Delta=-\cosh\lambda,\,\lambda\in(0,\infty).
\]
In this case the groundstate is known to be asymptotically double
degenerate, that is to say that the distance between the two lowest
eigenstates is exponentially small in the system size and there is
a finite gap to the third lowest state \citep{CG}.

The expression of the bulk energy
\begin{equation}
e_{\infty}(\Delta)=\frac{\cosh\lambda}{2}-\sinh^{2}\lambda\tilde{Y}(\lambda)\label{eq: e_bulk gapped}
\end{equation}
includes now an infinite sum
\begin{equation}
\tilde{Y}(\lambda)=\frac{1}{\sinh\lambda}\sum_{m\in\mathbb{Z}}\frac{\exp(-|m|\lambda)}{\cosh m\lambda}.\label{eq: Y^hat}
\end{equation}
It was shown in \citep{YY2} that the functions $Y(\gamma)$ and $\tilde{Y}(\lambda)$
(and hence the two expressions of $e_{\infty}(\Delta)$) can be thought
of as a continuation of one another from the real axis to the imaginary
one, $\lambda=\mathrm{i}\gamma$, such that all the derivatives are
continuous, while the difference is the function with essential singularity
at the origin.

The FSC in the gapped regime are exponentially small in the system
size, distinguishing however between the two asymptotically degenerate
groundstates. Their leading term
\begin{eqnarray}
E_{L}^{XXZ}(\Delta,e^{\mathrm{i}\varphi/L})-Le_{\infty}(\Delta)\label{eq: FSC gapped}\\
\,\,\,\,\,\,\,\,\,\,\,\,\,\,\,\,\simeq\mp\cos\left(\varphi/2\right)\sinh\lambda\frac{\sqrt{8k'}}{\pi^{3/2}\sqrt{L}}K(k)k_{1}^{L/2} & \times & \left(1+O(L^{-1})\right),\nonumber
\end{eqnarray}
is expressed in terms of the complete elliptic integral of the first
kind $K(k)$ with the elliptic modulus $k$ obtained as a solution
of $K(k')/K(k)=\lambda/\pi,$ the complementary modulus $k'=\sqrt{1-k^{2}}$
and the modulus
\[
k_{1}=\left(\frac{1-k'}{k}\right)^{2}=4e^{-\lambda}\prod_{n=0}^{\infty}\left[\frac{1+e^{-2\lambda(2n+2)}}{1+e^{-2\lambda(2n+1)}}\right]^{4},
\]
associated with the nome $e^{-2\lambda}.$ Note that $k_{1}<1$, when
$\lambda>0.$ The difference with the formula obtained in \citep{dVW}
is the only factor $\cos\left(\varphi/2\right)$ that incorporates all the dependence
on the twist angle. The real groundstate corresponds to the minus
sign, the two being non-crossing in the range $0<\varphi\leq\pi.$

There is also a scaling regime connecting the asymptotics (\ref{eq: FSC gapped})
and (\ref{eq:FSC crit}). It  corresponds to the scaling $\lambda\sim1/\ln L\to0$,
under which the FSC takes a scaling form
\begin{equation}
E_{L}^{XXZ}(\Delta,e^{\mathrm{i}\varphi/L})-Le_{\infty}(\Delta)=L^{-1}h(k'L,\varphi),\label{eq: scaling xxz}
\end{equation}
where $k'$ is the complementary modulus that vanishes in this limit
so that $k'L=O(1).$ An attempt to get the scaling function $h(x,\varphi)$,
such that $h(0,\varphi)$ is consistent with the $\gamma\to0$ limit
of (\ref{eq:FSC crit}), was undertaken in \citep{H2} for periodic
$(\varphi=0)$ and antiperiodic $(\varphi=\pi)$ boundary conditions.
The result obtained within an approximation reproduced the correct
limit for the periodic boundary conditions, while for the antiperiodic
ones it was not the case. We extended that calculation also for arbitrary
values of $\varphi$ and obtained a candidate for $h(x,\varphi)$,
such that the $\varphi$-dependent part of $h(0,\varphi)$ was twice
smaller than that of (\ref{eq:FSC crit}). This means that the approximation
used in \citep{H2} is not reliable for nonzero $\varphi$. The full
solution of this problem would require the analysis of FSC to the
density of Bethe roots in the spirit of \citep{HQB}. We leave it for
further investigation.

\subsection{Rate functions and cumulants }

Let us sketch the information about the statistics of $\mathcal{N}_{t}^{\lozenge}$
and $\mathcal{N}_{t}^{\circlearrowleft}$ that can be extracted from
knowing of the joint CGF $\hat{I}_{\mathcal{N}_{t}^{\circlearrowleft},\mathcal{N}_{t}^{\lozenge}}(\alpha,\beta).$
According to the previous discussion the latter can be obtained as
 the largest eigenvalue of the deformed operator $\mathcal{L}_{(\alpha,\beta)}$,
\[
\hat{I}_{\mathcal{N}_{t}^{\circlearrowleft},\mathcal{N}_{t}^{\lozenge}}(\alpha,\beta)=\Lambda_{0}(\alpha,\beta),
\]
which in turn is related to the groundstate eigenvalue of the twisted
XXZ chain (\ref{eq: Lambda_0(alpha,beta)}). We first consider each
of the two quantities $\mathcal{N}_{t}^{\lozenge}$ and $\mathcal{N}_{t}^{\circlearrowleft}$
separately and then discuss the most interesting features of their
mutual dependence.

\subsubsection{Large deviations of $\mathcal{N}_{t}^{\lozenge}$}

The scaled CGF
\begin{eqnarray*}
\hat{I}_{\mathcal{N}_{t}^{\lozenge}}(\beta)=\lim_{t\to\infty}t^{-1}\ln\mathbb{E}\exp\left(\beta\mathcal{N}_{t}^{\lozenge}\right) & = & \sum_{k=0}^{\infty}\frac{c_{k}(\mathcal{N}_{t}^{\lozenge})}{k!}\beta^{k}
\end{eqnarray*}
defines the scaled cumulants
\[
c_{k}(\mathcal{N}_{t}^{\lozenge})=\lim_{t\to\infty}t^{-1}\left\langle \left(\mathcal{N}_{t}^{\lozenge}\right)^{k}\right\rangle _{c},
\]
where $\left\langle \xi^{n}\right\rangle _{c}$ is the notation for
a usual cumulant of order $n$ of random variable $\xi$. According
to the above discussion it is given by $\hat{I}_{\mathcal{N}_{t}^{\lozenge}}(\beta)=\Lambda_{0}(0,\alpha)$
and in two leading orders as $L\to\infty$ is
\begin{eqnarray}
\hat{I}_{\mathcal{N}_{t}^{\lozenge}}(\beta) & = & L\left(\left(e^{\beta}-\frac{1}{4}e^{-\beta}\right)\,Y\left[\arccos(e^{-\beta}/2)\right]-1\right)\label{eq:rate tiles crit}\\
 & + & \frac{1}{L}\frac{\pi^{2}\sqrt{4e^{2\beta}-1}}{12\arccos(e^{-\beta}/2)}\left(1-\frac{2}{3}\frac{\pi}{\pi-\arccos(e^{-\beta}/2)}\right),\,\,\,\beta>-\ln2\nonumber
\end{eqnarray}
and
\begin{eqnarray}
\hat{I}_{\mathcal{N}_{t}^{\lozenge}}(\beta) & = & L\left(\left(\frac{1}{4}e^{-\beta}-e^{\beta}\right)\,\tilde{Y}\left[\mathrm{arccosh}(e^{-\beta}/2)\right]-1\right)\label{eq:rate tiles gapped}\\
 & + & \frac{a(\beta)}{\sqrt{L}}\exp(-L/\xi(\beta)),\,\,\,\beta<-\ln2,\nonumber
\end{eqnarray}
In the range $\beta<-\ln2$ we show explicitly only the dependence
of FSC on $L$, encapsulating the dependence on $\beta$ into two
functions $a(\beta),\xi(\beta)>0$. The definitions of these functions
in terms of elliptic integrals and moduli are clear from (\ref{eq: FSC gapped})
and are not important in context of present analysis.

\begin{figure}
\centering{}\includegraphics[width=0.4\paperwidth]{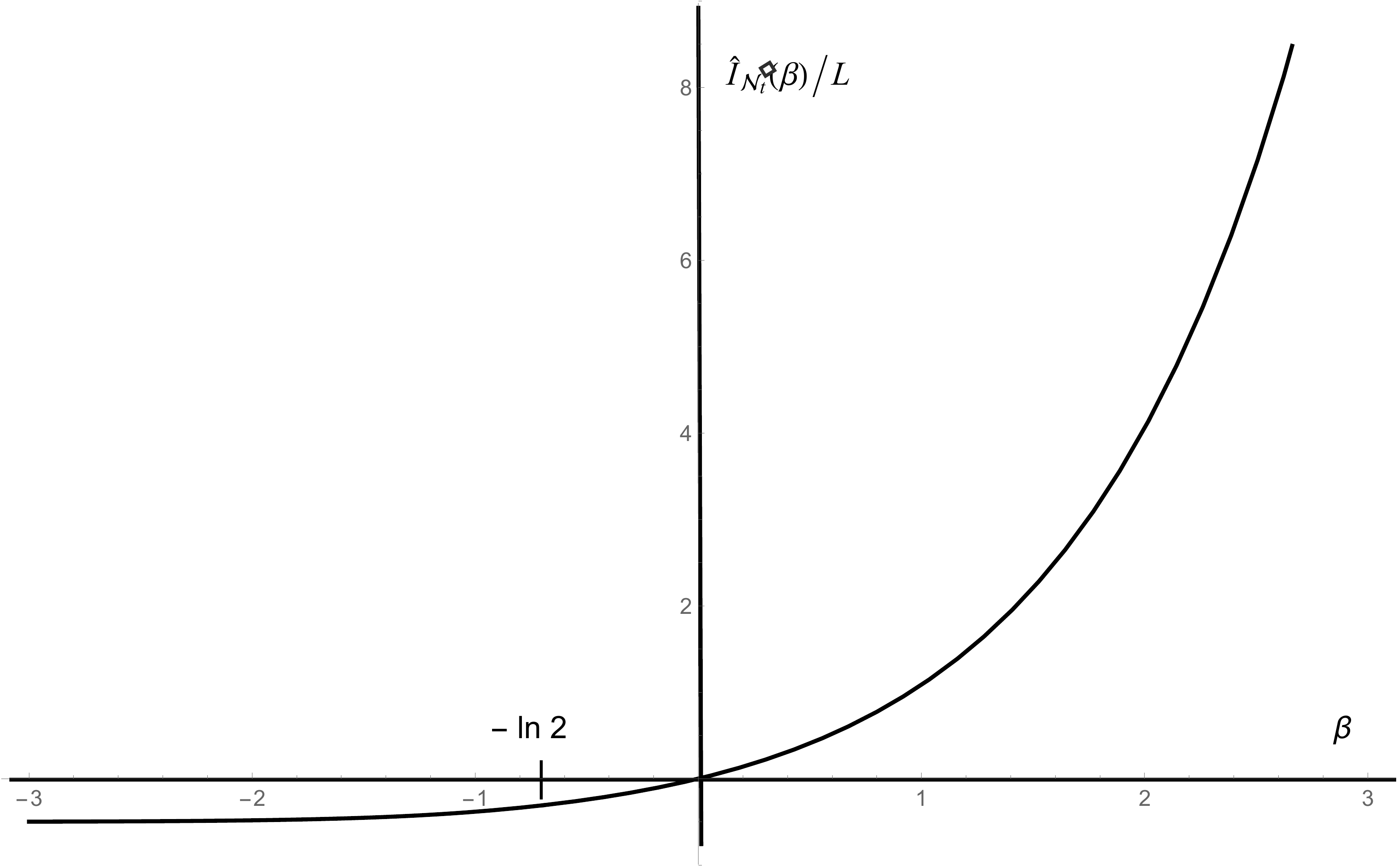}\caption{The bulk term of scaled cumulant generating function $\hat{I}_{\mathcal{N}_{t}^{\lozenge}}(\beta)/L$.
Two functions are sewn at $\beta=-\ln2$, so that the derivatives
of all orders are continuous.}\label{fig: I_bar}
\end{figure}

What we want to emphasize by this formulas is that  the scaled
CGF consists of two parts sewn at the point $\beta=-\ln2$. The bulk
part is smooth and convex everywhere and has the derivatives of all
orders continuous at this point  (see its plot in figure \ref{fig: I_bar}). In contrast, the FSC change drastically
there. They are of order of $1/L$ above $\beta=-\ln2$ and vanish
exponentially below with correlation length $\xi(\beta)$. The latter
is finite for $\beta<-\ln2$ and diverges as $\beta$ approaches the
upper bound of this domain.

To obtain the values of the scaled cumulants, among which the first
is the expected time average
\[
c_{1}(\mathcal{N}_{t}^{\lozenge})=\lim_{t\to\infty}\mathbb{E}\frac{\mathcal{N}_{t}^{\lozenge}}{t}
\]
and the second is the diffusion coefficient
\[
c_{2}(\mathcal{N}_{t}^{\lozenge})=\lim_{t\to\infty}\frac{\mathbb{E}\left(\mathcal{N}_{t}^{\lozenge}\right)^{2}-\left(\mathbb{E}\mathcal{N}_{t}^{\lozenge}\right)^{2}}{t},
\]
one needs to evaluate the derivatives of $\hat{I}_{\mathcal{N}_{t}^{\lozenge}}(\beta)$
at $\beta=0$. Technically nontrivial part is evaluating of derivatives
of the function $Y(x)$ from (\ref{eq: Y}) at $x=\pi/3$, which is
summarized in \ref{sec:Particular-values-of-Y}. Here we
show the first four cumulants of $\mathcal{N}_{t}^{\lozenge}$, evaluated
in two leading orders in $L$
\begin{eqnarray}
c_{1}(\mathcal{N}_{t}^{\lozenge}) & \simeq & \frac{5}{8}L-\frac{3}{8}L^{-1},\label{eq: c1 tile}\\
c_{2}(\mathcal{N}_{t}^{\lozenge}) & \simeq & \left(\frac{9\sqrt{3}}{2\pi}-\frac{11}{6}\right)L+\left(\frac{3\sqrt{3}}{8\pi}-\frac{1}{2}\right)L^{-1},\\
c_{3}(\mathcal{N}_{t}^{\lozenge}) & \simeq & \left(\frac{217}{32}-\frac{243}{4\pi^{2}}\right)L+\frac{81}{16\pi^{2}}L^{-1},\\
c_{4}(\mathcal{N}_{t}^{\lozenge}) & \simeq & \left(\frac{719}{12}-\frac{1701\sqrt{3}}{10\pi}+\frac{162}{\pi^{2}}+\frac{324\sqrt{3}}{\pi^{3}}\right)L+\left(\frac{135\sqrt{3}}{8\pi^{3}}+\frac{27}{2\pi^{2}}-2\right)L^{-1}.
\end{eqnarray}

Let us now look at the associated rate function $I_{\mathcal{N}_{t}^{\lozenge}}(y)$.
Since the scaled CGF is differentiable everywhere, the Gartner-Ellis
theorem provides the rate function being its Legandre transform. It
is defined parametrically by
\begin{eqnarray}
I_{\mathcal{N}_{t}^{\lozenge}}(y) & = & \beta\hat{I}'_{\mathcal{N}_{t}^{\lozenge}}(\beta)-\hat{I}_{\mathcal{N}_{t}^{\lozenge}}(\beta),\label{eq: rate legandre}\\
\phantom{I_{\mathcal{N}_{t}^{\lozenge}}(} y & = & \hat{I}'_{\mathcal{N}_{t}^{\lozenge}}(\beta),\label{eq: rate legandre 1}
\end{eqnarray}
where $\beta\in\mathbb{R}$ is supposed to be eliminated between two
expressions resulting in $I_{\mathcal{N}_{t}^{\lozenge}}(y)$ in the
domain $y>0$, and $I_{\mathcal{N}_{t}^{\lozenge}}(y)=\infty$ for
$y\leq0$.

\begin{figure}
\centering{}\includegraphics[width=0.5\textwidth]{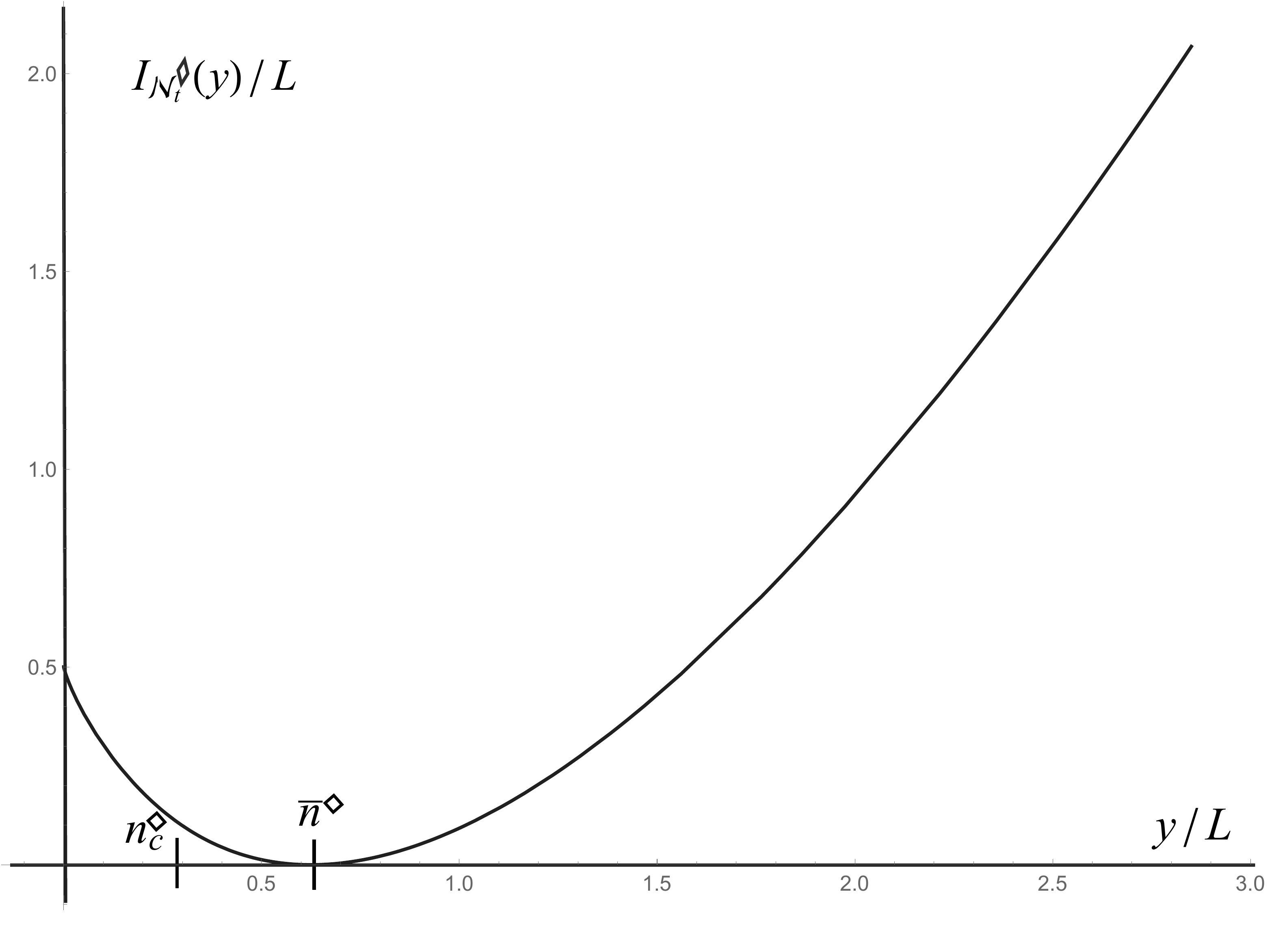}\caption{\label{fig: Rate function (tiles)}Rate function $I_{\mathcal{N}_{t}^{\lozenge}}(y)$.}
\end{figure}

It is clear that $\mathcal{N}_{t}^{\lozenge}/t$ and $I_{\mathcal{N}_{t}^{\lozenge}}(y)$
are extensive quantities, i.e. both are $O(L)$. Therefore, to rescale
them to finite bulk values we consider $I_{\mathcal{N}_{t}^{\lozenge}}(y)/L$
as a function of $y/L$, both supposed to be  finite as $L\to\infty$. The plot of this function is given in figure \ref{fig: Rate function (tiles)}.  Under
this scaling the function $I_{\mathcal{N}_{t}^{\lozenge}}(y)/L$,
has a single minimum at the point $y/L=\bar{n}^{\lozenge}=5/8$. The
critical point corresponding to $\beta=-\ln2$ is located at
\[
y=\hat{I}'_{\mathcal{N}_{t}^{\lozenge}}(-\ln2):=Ln_{c}^{\lozenge},
\]
where using the asymptotic form (\ref{eq:Y(x->0)}) of $Y(x)$ at
$x\to0$ we find
\[
n_{c}^{\lozenge}\simeq\frac{4\ln2-1}{6}\simeq0.2954.
\]
Like the scaled CGF the bulk part of the rate function is smooth everywhere,
while the FSC abruptly change the order of magnitude from $O(L^{-1})$
to $O(e^{-L/\xi(\beta)})$.

Using the asymptotic forms (\ref{eq:Y(x->Pi/2)},\ref{eq: Ytilde(x->infty)})
of $Y(x)$ and $\tilde{Y}(x)$ in the vicinity of $x=\pi/2$ and $x=\infty$,
respectively, we obtain the rate function in extremal regimes of small
and large values of $\mathcal{N}_{t}^{\lozenge}/t$.
\begin{eqnarray}
\frac{I_{\mathcal{N}_{t}^{\lozenge}}(y)}{L} & = & \frac{y}{L}\left(\ln\left(\frac{y\pi}{2L}\right)-1\right)-\left(\frac{3}{4}+\frac{1}{\pi^{2}}\right)+O(L/y),\,\,\,y/L\to\infty,\label{eq: rate function tiles y->infty}\\
\frac{I_{\mathcal{N}_{t}^{\lozenge}}(y)}{L} & = & \frac{1}{2}+\frac{y}{2L}\left(\ln\left(\frac{y}{2L}\right)-1\right)+O\left(\left(\frac{y}{L}\right)^{2}\right),\,\,\,y/L\to0.\label{eq: rate function tiles y->0}
\end{eqnarray}

\subsubsection{Large deviations of $\mathcal{N}_{t}^{\circlearrowleft}$}

\begin{figure}
\centering{}\includegraphics[width=0.5\textwidth]{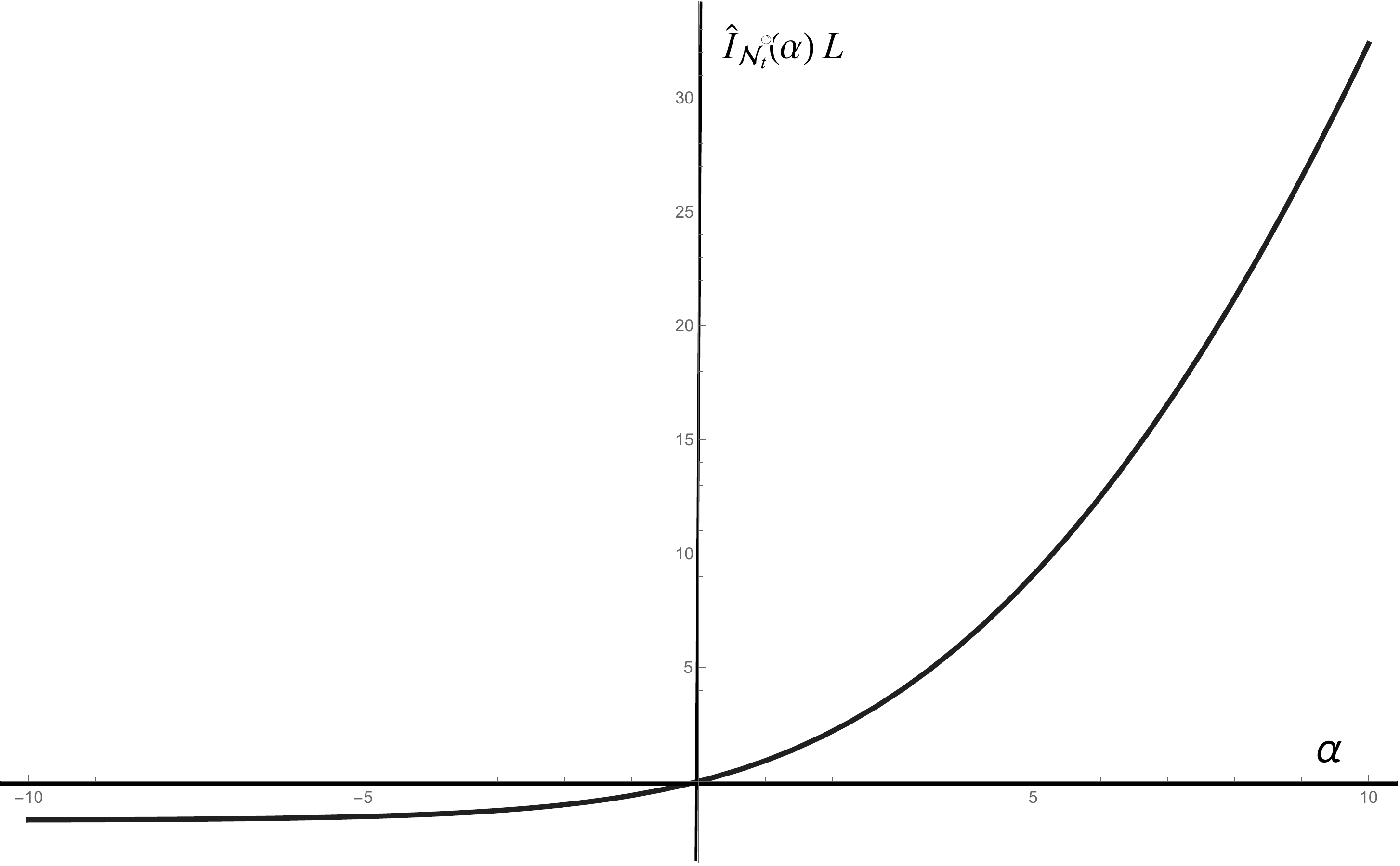}\caption{Scaled cumulant generating function of $\hat{I}{}_{\mathcal{N}_{t}^{\circlearrowleft}}(\beta)$ \label{fig: CGF global}}
\end{figure}

A similar set of data associated with the large deviations of $\mathcal{N}_{t}^{\circlearrowleft}$
is as follows. The leading order of the scaled CGF $\hat{I}{}_{\mathcal{N}_{t}^{\circlearrowleft}}(\alpha)=\Lambda_{0}(\alpha,0)$
is given by
\begin{equation}
\hat{I}{}_{\mathcal{N}_{t}^{\circlearrowleft}}(\alpha)=\frac{9\sqrt{3}}{4\pi L}\left(\left(\frac{\pi}{3}\right)^{2}-\left(\arccos\left[\frac{e^{\alpha/2}}{2}\right]\right)^{2}\right),\label{eq:CGF global}
\end{equation}
(see fig. \ref{fig: CGF global}). Therefore all the cumulants are of order of $L^{-1}.$

\begin{figure}
\centering{}\includegraphics[width=0.4\textwidth]{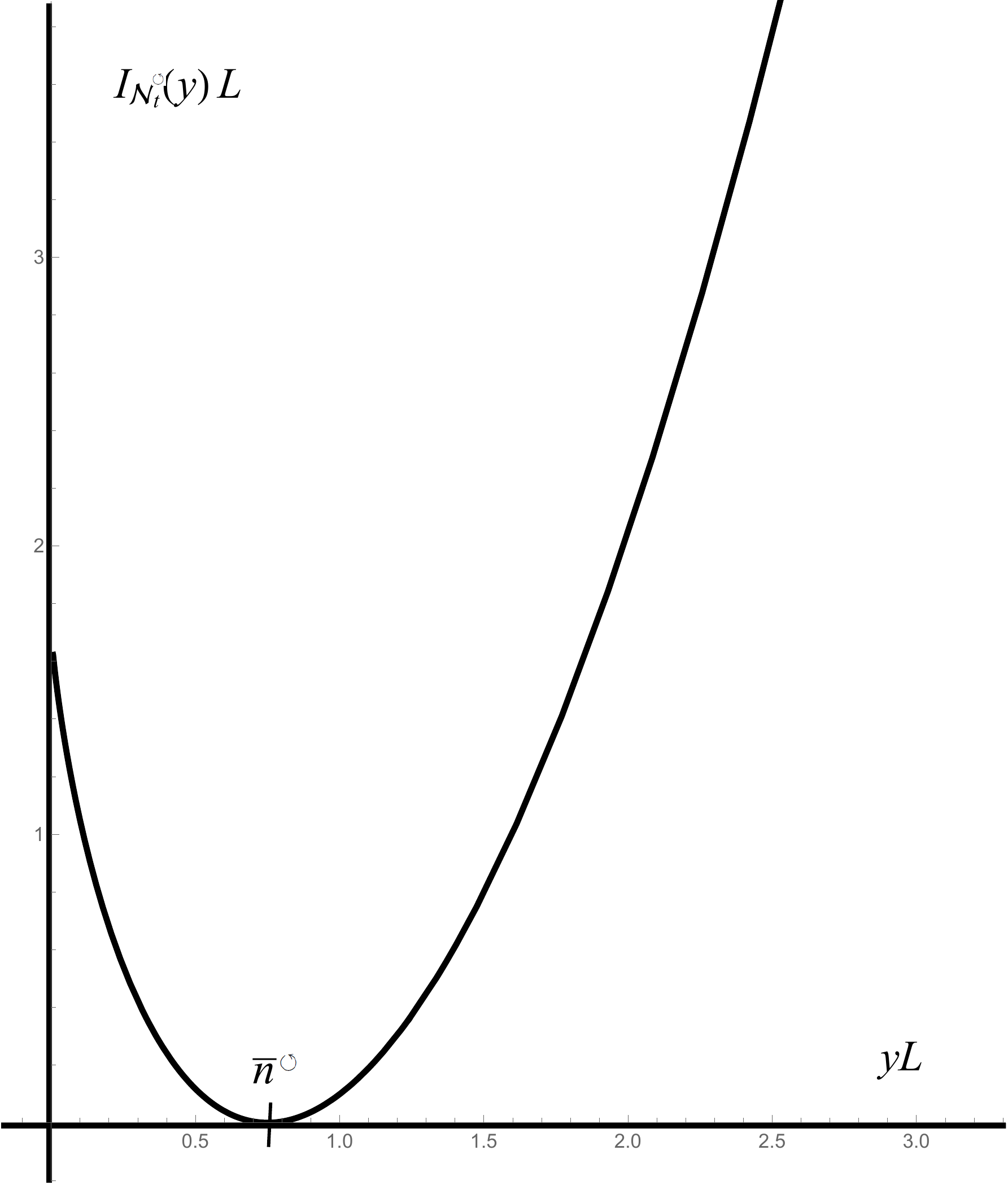}\caption{\label{fig:Rate-function (global)}Rate function $I_{\mathcal{N}_{t}^{\circlearrowleft}}(y)$}
\end{figure}

The first four cumulants are
\begin{eqnarray}
c_{1}(\mathcal{N}_{t}^{\circlearrowleft}) & \simeq & \frac{3}{4}\frac{1}{L},\label{approve2}\\
c_{2}(\mathcal{N}_{t}^{\circlearrowleft}) & \simeq & \left(\frac{1}{2}-\frac{3\sqrt{3}}{8\pi}\right)\frac{1}{L},\\
c_{3}(\mathcal{N}_{t}^{\circlearrowleft}) & \simeq & \left(\frac{1}{2}-\frac{3\sqrt{3}}{4\pi}\right)\frac{1}{L},\\
c_{4}(\mathcal{N}_{t}^{\circlearrowleft}) & \simeq & \left(\frac{5}{6}-\frac{3\sqrt{3}}{2\pi}\right)\frac{1}{L}.
\end{eqnarray}
The rate function $I_{\mathcal{N}_{t}^{\circlearrowleft}}(y)$ is
again obtained from the scaled CGF as the Legendre transform. Its
plot is given in fig.~\ref{fig:Rate-function (global)}.

It is defined parametrically via formulas similar to (\ref{eq: rate legandre},\ref{eq: rate legandre 1})
up to change the subscript to $\mathcal{N}_{t}^{\circlearrowleft}$.
It follows then that the magnitude of the rate function is of order
of $I_{\mathcal{N}_{t}^{\circlearrowleft}}(y)=O(1/L)$, while expressions
obtained are applicable in the limit $y\to0,L\to\infty,yL=const$. Then
$LI_{\mathcal{N}_{t}^{\circlearrowleft}}(y)$ is the finite function
of $Ly$ that has a single minimum at $Ly=\bar{n}^{\circlearrowleft}=3/4$.
The asymptotic behaviour of the $LI_{\mathcal{N}_{t}^{\circlearrowleft}}(y)$
corresponding to the limits $\text{\ensuremath{\alpha\to\pm\infty}}$.
\begin{eqnarray*}
I_{\mathcal{N}_{t}^{\circlearrowleft}}(y)L & = & \frac{5\pi\sqrt{3}}{16}+Ly\left(2\ln(Ly)-2+\ln\frac{2^{8}}{3^{5}}\right)+O((Ly)^{2}),yL\to0\\
I_{\mathcal{N}_{t}^{\circlearrowleft}}(y)L & = & \frac{4\pi\left(Ly\right)^{2}}{9\sqrt{3}}-\frac{\sqrt{3}\pi}{4}+O(e^{-Ly}),\,\,\,yL\to\infty.
\end{eqnarray*}

\subsubsection{Mutual dependence of $\mathcal{N}_{t}^{\lozenge}$ and $\mathcal{N}_{t}^{\circlearrowleft}$}

The simplest characteristics of mutual dependence of two random variables
is the covariance. In our case this is the scaled covariance, the
mixed scaled cumulant of the first degree in each variable,
\begin{eqnarray*}
c_{1,1}\left(\mathcal{N}_{t}^{\circlearrowleft},\mathcal{N}_{t}^{\lozenge}\right) & = & \lim_{t\to\infty}t^{-1}\left(\mathbb{E}\left(\mathcal{N}_{t}^{\circlearrowleft}\mathcal{N}_{t}^{\lozenge}\right)-\mathbb{E}\mathcal{N}_{t}^{\circlearrowleft}\mathbb{E}\mathcal{N}_{t}^{\lozenge}\right)\\
 & = & \left.\frac{\partial^{2}\hat{I}_{\mathcal{N}_{t}^{\circlearrowleft},\mathcal{N}_{t}^{\lozenge}}(\alpha,\beta)}{\partial\alpha\partial\beta}\right|_{\alpha,\beta=0}.
\end{eqnarray*}
Evaluating the derivatives of $\Lambda_{0}(\alpha,\beta)$ we obtain
\[
Lc_{1,1}\left(\mathcal{N}_{t}^{\circlearrowleft},\mathcal{N}_{t}^{\lozenge}\right)=1-\frac{3\sqrt{3}}{8\pi}\simeq0.793252.
\]
The more detailed characteristics of mutual dependence that gives
a meaning to the transition between the gapless and the gapped phases
in the groundstate of the XXZ model is the conditional scaled CGF
and conditional rate function. Specifically let us consider the scaled
CGF of $\mathcal{N}_{t}^{\circlearrowleft}$ conditioned to the value
of $\mathcal{N}_{t}^{\lozenge}$
\[
\hat{I}_{\mathcal{N}_{t}^{\circlearrowleft}|\mathcal{N}_{t}^{\lozenge}}(\alpha|y)=\lim_{t\to\infty}t^{-1}\ln\mathbb{E}\left(e^{\alpha\mathcal{N}_{t}^{\circlearrowleft}}|\mathcal{N}_{t}^{\lozenge}=y\right).
\]
Taking into account the form of the eigenvalue $\Lambda_{0}(\alpha,\beta)=\Lambda_{0}^{0}(\beta)+\Lambda_{0}^{1}(\alpha,\beta)$,
where $\Lambda_{0}^{0}(\beta)=O(L)$ is the bulk part dependent only
on $\beta$ and $\Lambda_{0}^{1}(\alpha,\beta)=o(L)$ is the FSC,
we have
\begin{eqnarray*}
\hat{I}_{\mathcal{N}_{t}^{\circlearrowleft}|\mathcal{N}_{t}^{\lozenge}}(\alpha|y) & \simeq & \Lambda_{0}^{1}(\alpha,\beta(y))-\Lambda_{0}^{1}(0,\beta(y)),
\end{eqnarray*}
where $\beta(y)$ is the solution of
\[
y=\frac{\partial\Lambda_{0}^{0}(\beta)}{\partial\beta}.
\]
The explicit formulas depend on the value of $y$. For $y/L\geq n_{c}^{\lozenge}$
that corresponds to $\beta(y)>-\ln2$ we have the conditioned scaled
CGF, which coincides with the unconditioned one (\ref{eq:CGF global})
up to an overall $y$-dependent factor

\[
\hat{I}_{\mathcal{N}_{t}^{\circlearrowleft}|\mathcal{N}_{t}^{\lozenge}}(\alpha|y)=\frac{g(\beta(y))}{L}\left(\left(\frac{\pi}{3}\right)^{2}-\left(\arccos\left[\frac{e^{\alpha/2}}{2}\right]\right)^{2}\right).
\]
Here
\[
g(\beta)=\frac{\pi\sqrt{4e^{2\beta}-1}}{2\arccos\frac{e^{-\beta}}{2}\left(\pi-\arccos\frac{e^{-\beta}}{2}\right)},
\]
$\beta(y)$ solves the equation

\[
y/L=\frac{d}{d\beta}\left(\left(e^{\beta}-\frac{1}{4}e^{-\beta}\right)\,Y\left[\arccos(e^{-\beta}/2)\right]\right),
\]
and we imply that $y=O(L)$.

For $y/L<n_{c}^{\lozenge}$ we obtain purely exponential function
of $\alpha$ scaled by the $y$-dependent factor that decays exponentially
as $L$ grows to infinity.
\[
\hat{I}_{\mathcal{N}_{t}^{\circlearrowleft}|\mathcal{N}_{t}^{\lozenge}}(\alpha|y)=\left(e^{\alpha/2}-1\right)a(\beta(y))e^{-L/\xi(\beta(y))}L^{-1/2}.
\]
Here, $\beta(y)$ solves the equation
\[
y/L=\frac{d}{d\beta}\left(\left(\frac{1}{4}e^{-\beta}-e^{\beta}\right)\,\tilde{Y}\left[\mathrm{arccosh}(e^{-\beta}/2)\right]-1\right)
\]
and the functions $a(\beta),\xi(\beta)$ are those appeared in (\ref{eq:rate tiles gapped}).

Thus, up to the  overall scale factor, which can be incorporated
into the time scale, the scaled CGF of $\mathcal{N}_{t}^{\circlearrowleft}$
conditioned to a value of $\mathcal{N}_{t}^{\lozenge}$ has two different
functional forms depending on whether the $\mathcal{N}_{t}^{\lozenge}/(Lt$)
is greater or less than $n_{c}^{\lozenge}$. This sharp change of
behaviour at the critical point can be made continuous upon looking
at it in the smaller scale. The scaling regime of the groundstate
of the XXZ Hamiltonian mentioned in the end of subsection \ref{subsec:The-groundstate-of XXZ}
suggests that if we consider the scaling
\[
y/L=n_{c}^{\lozenge}-\frac{s}{\log L}
\]
with $s>0$, the conditional scaled CGF will take the form
\[
\hat{I}_{\mathcal{N}_{t}^{\circlearrowleft}|\mathcal{N}_{t}^{\lozenge}}(\alpha|y)=L^{-1}f(s,\alpha),
\]
where $f(s,\alpha)$ is a crossover function related to $h(x,\varphi)$
from (\ref{eq: scaling xxz}) by a suitable variable change. The two
functions obtained above are expected to be restored as limiting cases
of $f(s,\alpha)$ in the limits $s\to0$ and $s\to\infty.$ An explicit
form of this function is yet to be determined.

\section{Discussion and conclusion. \label{sec: discussion}}

To summarize, we obtained the large deviation functions for two additive
functionals on the trajectories of RPM: the total number of tiles removed
by avalanches $\mathcal{N}_{t}^{\lozenge}$ and the total number of
global avalanches $\mathcal{N}_{t}^{\circlearrowleft}.$ Technically,
all the properties of the large deviations described stem from the
corresponding properties of the groundstate of twisted XXZ Hamiltonian.
What we would like to do here is to provide them with a stochastic
interpretation.

First, we focused on the expressions of the scaled CGM $\hat{I}_{\mathcal{N}_{t}^{\lozenge}}(\beta)$
and $\hat{I}_{\mathcal{N}_{t}^{\circlearrowleft}}(\alpha)$ separately
and on the corresponding rate functions $I_{\mathcal{N}_{t}^{\lozenge}}(x),I_{\mathcal{N}_{t}^{\circlearrowleft}}(y)$
in the limit $L\to\infty$. The first outcome of these results are
the asymptotics of the scaled cumulants of $\mathcal{N}_{t}^{\lozenge}$
in two leading orders and of $\mathcal{N}_{t}^{\circlearrowleft}$
in the leading order in $L$. The cumulants of the first order, the
means, can be also obtained from an averaging over the stationary
state. The asymptotic formulas (\ref{eq: c1 tile},\ref{approve2})
obtained indeed reproduce the large $L$ expansions of the conjectured
exact formulas (\ref{eq: mean total},\ref{eq: mean global}) respectively.
The cumulants of higher orders involve unequal time correlations.
Thus, they are beyond the scope of the stationary state analysis and
to our knowledge are presented  for the first time. To find exact
expression of the cumulants at arbitrary finite $L$ one can try the
technique of perturbative solution of T-Q relations which was successfully
applied to the asymmetric simple exclusion process in \citep{PM}.
The case of RPM however seems even more challenging because of
nontrivial structure of the groundstate solution of Bethe equations
even at the stochastic point.

Our analysis also explains the observation made in \citep{AR-13} of
the spin current in the XXZ chain with twist $\varphi=-2\pi/3$ being
related to the particle current in nonlocal asymmetric simple exclusion process (NASEP) via the factor $\sqrt{3}$. The NASEP was proposed in   \citep{AR-13} as a model obtained from RPM by the usual interface-particle system correspondence.
Specifically, the upper boundary of a stable configuration
in the periodic Dyck path representation is associated to an $L/2$-particle configuration
on a one-dimensional lattice consisting of $L$ sites. A down-step
of the line going from horizontal position $i$ to $i+1$ is mapped
to the site $i$ of the 1D lattice occupied by a particle and the
up-step is mapped to an empty site. The time evolution of the RPM is also  naturally
mapped to the evolution of the particle system.

In particular, the NASEP current per bond is defined in \citep{AR-13} as the
number of right jumps minus the number of left jumps of particles
crossing a bond per unit time. Alternatively one can think of the
total distance traveled by particles, i.e. difference of numbers of
right and left jumps on the whole lattice, divided by $L$. Every
adsorption of a tile in a local minimum (valley) contribute $+1$
and the avalanche of size $n$ contribute $-(n-1)$ into the into
the distance. Here, minus one corresponds to the extra tile starting
the avalanche that was included into the avalanche size and does not
contribute to the distance traveled by particles  in NASEP. The
global avalanches do not contribute to the current according to definition
of \citep{AR-13}. As the number of adsorbed tiles should be approximately
(up to a bounded difference) equal to the number of tiles removed
by both local and global avalanches, the total distance traveled by
all the particles in NASEP approximately equals $-\mathcal{N}_{t}^{\circlearrowleft}$.
Thus, up to the sign the mean current $J$ in NASEP coincides with
$c_{1}(\mathcal{N}_{t}^{\circlearrowleft})$.

At the same time the spin current in $XXZ$ model defined as $J^{z}=\left\langle 0\left|J_{i}^{z}\right|0\right\rangle ,$
where $J_{i}^{z}=\mathrm{i}(\sigma_{i}^{+}\sigma_{i+1}^{-}-\sigma_{i}^{-}\sigma_{i+1}^{+})$
and the averaging is over the groundstate of the XXZ Hamiltonian,
can be evaluated as $J^{z}\simeq-\partial E_{L}^{XXZ}(\Delta,e^{\mathrm{i}\varphi/L})/\partial\varphi$
in the leading order in $L$. Going from the differentiation of the
XXZ energy in the twist variable $\varphi$ to that of the eigenvalue
of the stochastic generator $\mathcal{L}_{(\alpha,\beta)}$ in the
variable $\beta$ we obtain the desired relation.

Let us discuss the large deviation functions beyond the cumulants.
The quantity $\mathcal{N}_{t}^{\lozenge}$ grows by elementary random
increments $\mathcal{\delta N}_{\tau}^{\lozenge}(x)$ associated with
space-time positions in $\left\{ (x,\tau)\in(1,\dots,L)\times[0,t]\right\} $.
The total rate of growth is of order of $L$, which suggests that the
main part of the scaled CGF $\hat{I}_{\mathcal{N}_{t}^{\lozenge}}(x)$
scales linearly with $L$ and, hence, the usual CGF of $\mathcal{N}_{t}^{\lozenge}$
is linear in $Lt$. If the increments were uncorrelated, the CGF would
be purely linear. The spacial correlations lead to appearance of FSC
to the scaled CGF.

Usually, the FSC to the bulk free energy in equilibrium systems come
from the geometric factors like surface, edges, corners e.t.c., each
contribution being scaled as an integer power of system size according
to its dimension. These contributions, however, are not present in
the systems with periodic boundary conditions \citep{FB}. In such
systems away from criticality, which are characterized by  a finite
correlation length, the FSC are exponentially small. In critical systems
with infinite correlation length there is also the fluctuation induced
Casimir-like term  that decays as a power of system size
\citep{FB}. Thinking about CGM as an analogue of the free energy
one can expect similar scenario in the non-equilibrium setting. Our
analysis of $\hat{I}_{\mathcal{N}_{t}^{\lozenge}}(\beta)$ reveals
the transition from critical phase, $\beta>-\log2$, with the FSC
of order of $O(1/L)$ to the noncritical phase, $\beta<-\log2$, where
the FSC decay exponentially.

In particular the system is critical at $\beta=0$. An indication
of the infinite correlation length is the presence of global avalanches,
which happen with the frequency of order of inverse system size. The
probability distribution of the number of global avalanches per unit
time behaves as
\begin{equation}
\mathbb{P}\left(\frac{\mathcal{N}_{t}^{\circlearrowleft}}{t/L}\approx x\right)\asymp\exp\left(-\frac{t}{L}I(x)\right),\label{eq:global distrib}
\end{equation}
where we used the notation $I(x)=LI_{\mathcal{N}_{t}^{\circlearrowleft}}(x)$
to remove the $L$-dependence from the rate function and to emphasize
that the effective time units scale linearly with system size. This
time scaling suggests the dynamical exponent to be $z=1$, which is
the usual companion for conformal invariance. The universality of
global avalanche statistics over the critical phase and the phase
transition to the non-critical phase are most clearly manifested with
the rate function for $\mathcal{N}_{t}^{\circlearrowleft}$ conditioned
to a given value of $\mathcal{N}_{t}^{\lozenge}/t=yL.$ To this end,
it is convenient to introduce the effective unit of time $\tau(y)$
being the coefficient of $\hat{I}_{\mathcal{N}_{t}^{\circlearrowleft}|\mathcal{N}_{t}^{\lozenge}}(\alpha|y)$
encapsulating all the dependence on $y$. $\tau(y)$ scales linearly
with $L$, $\tau(y)=O(L)$, in the critical phase $y>n_{c}^{\lozenge}$
and grows exponentially, $\tau(y)=O(e^{L/\xi(\beta(y))}\sqrt{L})$
otherwise, i.e. $0<y<n_{c}^{\lozenge}$. Then the conditioned probability
can be given in a universal form
\[
\mathbb{P}\left(\left.\frac{\mathcal{N}_{t}^{\circlearrowleft}}{(t/\tau(y))}\approx x\right|\frac{\mathcal{N}_{t}^{\lozenge}}{t}=yL\right)\asymp\exp\left(-\frac{t}{\tau(y)}I(x)\right),
\]
where $I(x)$ is the same as in (\ref{eq:global distrib}) over the
whole critical phase and changes to
\[
I(x)=2x(\log2x-1)+1
\]
 in the non-critical one. The latter is nothing but the rate function
of the Poisson distribution, which suggests the global avalanches
being independent events, i.e. they happen so rare in the non-critical
regime that the system forgets the past from one global avalanche
to another.

We also explicitly obtained the tails of the rate functions. Sometimes
from the tails one can get an idea how the system should be modified
to fall into an extremal regime. In particular we found that $I_{\mathcal{N}_{t}^{\lozenge}}(y)\to1/2$
as $y\to0$, which means that the probability of trajectories keeping
$\mathcal{N}_{t}^{\lozenge}$ small behaves as $\exp\left(-Lt/2\right)$.
We interpret this regime as the system remaining in the substrate
configuration by forbidding for the tiles to arrive anywhere except
the peaks, i.e. at $L/2$ sites. The rate functions $I_{\mathcal{N}_{t}^{\circlearrowleft}}(x)$
also tends to a constant as $x\to0.$ As $x\to\infty$ the functions
$I_{\mathcal{N}_{t}^{\lozenge}}(x)$ and $I_{\mathcal{N}_{t}^{\circlearrowleft}}(x)$
behave in Poissonian-like and Gaussian-like manner respectively. The
interpretation of the three latter regimes is yet to be found. In
general detailed studies of the stochastic processes conditioned to
atypical fluctuations requires studies of the maximal eigenvector
of the stochastic generator rather than the eigenvalue. This is the subject
for further work.

As another unsolved problems we leave  the construction of the
crossover function, which would connect the regimes $\mathcal{N}_{t}^{\lozenge}/t>n_{c}^{\lozenge}L$
and $\mathcal{N}_{t}^{\lozenge}/t<n_{c}^{\lozenge}L$ of the conditional
rate function $I_{\mathcal{N}_{t}^{\circlearrowleft}|\mathcal{N}_{t}^{\lozenge}}(\alpha|y)$.

\ack{}{}

The work is supported by Russian Foundation of Basic Research under
grant 17-51-12001, DFG grant RI 317/17-1, Heiseberg-Landau
Program and Russian Academic Excellence Project '5-100'.

\appendix

\section{Particular values and asymptotics of $Y(x)$ and $\tilde{Y}(x)$.
\label{sec:Particular-values-of-Y}}

Here we present the result of explicit evaluation of integral (\ref{eq: Y})
in $Y(x)$ and its derivatives at points $x=\pi/3,0,\pi/2$. It can
be done using the residue calculus by closing the contour of integration
through the upper (lower) half-planes. This becomes possible after
introducing a suppressing factor $e^{i\epsilon x}(e^{-i\epsilon x})$
with small $\epsilon>0$ into the integrand and sending $\epsilon$
to zero in the end. The values of derivatives of $Y^{(n)}(\pi/3)$
are listed below up to $n=6$.
\begin{eqnarray}
Y\left(\frac{\pi}{3}\right) & = & \frac{4}{3},\\
Y'\left(\frac{\pi}{3}\right) & = & -\frac{25}{6\sqrt{3}},\\
Y^{''}\left(\frac{\pi}{3}\right) & = & \frac{18\sqrt{3}}{\pi}-3,\\
Y^{(3)}\left(\frac{\pi}{3}\right) & = & \frac{463}{8\sqrt{3}}-\frac{54}{\pi}-\frac{243\sqrt{3}}{\pi^{2}},\\
Y^{(4)}\left(\frac{\pi}{3}\right) & = & \frac{15059}{18}-\frac{9306\sqrt{3}}{5\pi}+\frac{972}{\pi^{2}}+\frac{3888\sqrt{3}}{\pi^{3}},\\
Y^{(5)}\left(\frac{\pi}{3}\right) & = & -\frac{33185\sqrt{3}}{4}+\frac{8946}{\pi}+\frac{72495\sqrt{3}}{\pi^{2}}-\frac{19440}{\pi^{3}}-\frac{72900\sqrt{3}}{\pi^{4}},\\
Y^{(6)}\left(\frac{\pi}{3}\right) & = & -\frac{3938533}{6}+\frac{10658034\sqrt{3}}{7\pi}-\frac{425250}{\pi^{2}}-\frac{2658420\sqrt{3}}{\pi^{3}},\\
 &  & +\frac{437400}{\pi^{4}}+\frac{1574640\sqrt{3}}{\pi^{5}}.
\end{eqnarray}
The expansion around $x=0$ and $x=\pi/2$ yields
\begin{eqnarray}
Y(x) & = & \frac{2\ln2}{x^{2}}+\frac{\ln2}{3}-\frac{1}{6}+O(x)\text{,\,\,}x\to0,\label{eq:Y(x->0)}\\
Y\left(\frac{\pi}{2}+x\right) & = & \frac{2}{\pi}-\left(\frac{1}{2}+\frac{2}{\pi^{2}}\right)x+\frac{24+17\pi^{2}}{9\pi^{3}}x^{2}+O(x^{3}),\,\,\,x\to0.\label{eq:Y(x->Pi/2)}
\end{eqnarray}
Finally we provide  the expansion of the sum in (\ref{eq: Y^hat})
\begin{equation}
\tilde{Y}(x)=2e^{-x}+10e^{-3x}+10e^{-5x}+O(e^{-7x}),\,\,\,x\to\infty.\label{eq: Ytilde(x->infty)}
\end{equation}

\end{document}